\documentclass[journal,12pt,draftclsnofoot,onecolumn]{IEEEtran}

\normalsize

\usepackage{graphicx}
\usepackage{epstopdf}

 \DeclareGraphicsExtensions{.pdf}

\usepackage[cmex10]{amsmath}
\usepackage{bm}
\usepackage{epsfig}
\usepackage{booktabs} 
\usepackage{latexsym}
\usepackage{enumerate}
\usepackage{multirow}
\usepackage{epstopdf}
\usepackage{longtable}
\usepackage{stfloats}
\usepackage{color} 
\usepackage{amssymb}
\graphicspath{{./Figures/}}
\usepackage{color}
\usepackage{tikz}
\usepackage{bbm}
\usepackage{authblk}
\usepackage{verbatim}
\usepackage{mathtools}

\usepackage[tight,footnotesize]{subfigure}
\usepackage{balance}

\usepackage{siunitx}
\DeclareSIUnit\bps{bps}
\DeclareSIUnit\Torr{Torr}
\DeclareSIUnit\torr{Torr}
\DeclareSIUnit\sample{Sa}
\usepackage{cite}

\usepackage{color, soul}
\newcommand{\tabincell}[2]{\begin{tabular}{@{}#1@{}}#2\end{tabular}}

\begin{document}

\title{Channel Measurement, Characterization and Modeling for Terahertz Indoor Communications Above 200 GHz}

\author{Yi~Chen,~\IEEEmembership{Student~Member,~IEEE}, Chong~Han,~\IEEEmembership{Member,~IEEE}, Ziming~Yu and Guangjian~Wang,~\IEEEmembership{Member,~IEEE}

\thanks{
This paper was presented in part at the Fourth International Workshop on Terahertz Communications, in conjunction with IEEE ICC 2021~\cite{he2021channel}. 
}
\thanks{Y. Chen and C. Han are with the Terahertz Wireless Communications Laboratory, Shanghai Jiao Tong University, Shanghai, China (email: \{yidoucy, chong.han\}@sjtu.edu.cn).
\par Z. Yu and G. Wang are with Huawei Technologies Co., Ltd, Chengdu, China (email: \{yuziming,wangguangjian\}@huawei.com).}
}
\markboth{}{Chen
\MakeLowercase{\textit{et al.}}:201-209 GHz Channel Measurement, Characterization and Modeling for Terahertz Indoor Communications} 
\maketitle 
\thispagestyle{empty}

\begin{abstract}
Terahertz (THz) communications are envisioned as a promising technology for sixth-generation (6G) and beyond systems, owing to its unprecedented multi-gigahertz (GHz) bandwidth. In this paper, channel measurement campaigns in indoor scenarios at 201-209~GHz are reported. Four different communication scenarios including 90 transmitter-receiver pairs are measured in two channel measurement campaigns of a meeting room and an office room, respectively. The two measurement campaigns contains four scenarios, namely, a meeting room, cubicle area, hallway and non-line-of-sight (NLoS) case. The propagation of multi-path components (MPCs) in the four scenarios is characterized by the power-delay-angular profiles. Based on them, the temporal and spatial consistency for varying receiver locations in the complex hallway and NLoS scenarios are verified. To characterize, the large-scale best-direction and omni-directional path losses in indoor scenarios are separately analyzed and modeled by the close-in (CI) model. Furthermore, the small-scale channel parameters, e.g., the number of clusters, delay spread, angular spread, and cluster time-of-arrival are analyzed and modeled by proper distributions. As a general framework, a ray-tracing-statistical hybrid model is proposed for wireless propagation at 201-209~GHz, although, admittedly, the measurement results and analysis reveal that the channel characteristics in various indoor scenarios exhibit noticeable differences that need tailored parameter settings.
\boldmath

\end{abstract}
\section{Introduction}
New spectral bands are required to support Terabit-per-second (Tbps) data rates for future wireless applications to accommodate the exponential growth of wireless data traffic~\cite{chong2017thz,rappaport2019wireless,akyildiz2022terahertz }. Currently, wireless local area networks (WLAN) techniques, i.e., 802.11ad protocol and fifth-generation (5G) mobile networks, have opened up the millimeter-wave (mmWave) spectrum (10-100~GHz) to seek for broader bandwidth and higher data rates. However, the mmWave spectrum is still limited to several-GHz bandwidth and thus cannot support Tbps rates. To further move up with the carrier frequency, the Terahertz (THz) band spanning over 0.1 and 10~THz, is envisioned as one of the promising spectrum bands to enable ultra-broadband sixth-generation (6G) communications~\cite{wang20206G}.

\par Investigation of wireless channels is the foundation for designing wireless communication systems in the new spectrum band, which includes the study of wave propagation, development of channel models and understanding of characteristic channel properties. Channel measurement realized by channel sounders is the fundamental of channel studies, since physical measurement results are essential both to discover new channel characteristics or to uphold modeling and simulation results. Current THz channel measurement systems are typically based on three approaches, namely, vector network analyzer (VNA), sliding correlation, and THz time-domain spectroscopy (THz-TDS) channel study~\cite{wang2022channelsurvey}.
From the literature, a number of channel measurement campaigns at THz frequencies have been reported for short-range ($<10$~m) indoor or room-scale scenarios~\cite{yu2020wideband,yi2021Channel,xing2018propagation,xing2019indoor,kim2016characterization,eckhardt2019measurements,cheng2020thz,fu2020modeling,abbasi2020channel,song2020channel,ju2021millimeter,zantah2019channel}, long-range (10-100~m) indoor scenarios~\cite{Nguyen2018Comparing}, outdoor vehicular communications~\cite{guan2021channel,petrov2020measurements,eckhardt2021channel} and long-range urban scenarios~\cite{abbasi2019double,he2021wireless}, elaborated as follows.

Limited by the high path loss and frequency-dependent molecular absorption in THz wireless channels, to-date measurement campaigns of THz channels mainly focus on short-range scenarios, e.g., on a table, on a computer motherboard, and inter-racks in data centers, with the distance ranging from 0.1~m to 10~m~\cite{priebe2011channel,kim2016characterization,song2020channel,eckhardt2019measurements,cheng2020thz,fu2020modeling,xing2019indoor}.
For instance, \textit{Priebe~et~al.} studied the desktop scenario at 300~GHz in \cite{priebe2011channel}. Moreover, \textit{Kim~et~al.} measured the THz propagation in the 300-320~GHz frequency band in the computer motherboard environment including line-of-sight (LoS), reflected non-line-of-sight (RNLoS), obstructed line-of-sight (OLoS) and non-line-of-sight (NLoS) scenarios~\cite{kim2016characterization}. Besides, \textit{Cheng~et~al.} measured the data center environment at 300-320~GHz, including seven scenarios simulating the transmitter/receiver (Tx/Rx) misalignment and cable obstruction, where the path loss and multipath components (MPCs) are analyzed~\cite{cheng2020thz}.
\par By contrast, other indoor channel measurement campaigns conduct room-scale studies, e.g., in lecture rooms, office rooms and meeting rooms~\cite{priebe2011channel,xing2018propagation,Nguyen2018Comparing,ju2021millimeter,yu2020wideband,yi2021Channel,he2021channel,chen2021channel}. 
For example, \textit{Priebe~et~al.} scanned a small office room at multiple pairs of angles of arrival (AoA) and angles of departure (AoD) for detectable paths in the 300~GHz band, with the transmitter-receiver (Tx-Rx) distance of about 1.67~m~\cite{priebe2011channel}.
However, indoor channel measurement campaigns that focus on the frequency band higher than 300~GHz typically suffer from a measuring distance smaller than 5~m. In order to study indoor channels in practical room-scale scenarios, researchers typically turn to lower frequency bands to achieve larger measuring distances~\cite{abbasi2020channel,Nguyen2018Comparing}.
For instance, in our previous works~\cite{chen2021channel,yu2020wideband}, a typical indoor meeting room is measured at 140~GHz with the Tx-Rx distance ranging from 1.8~m to 7.3~m. Besides, \textit{Nguyen~et~al.} measured the indoor channels in a shopping mall at 140~GHz with measurement distance from 3~m to 65~m~\cite{Nguyen2018Comparing}.
Nevertheless, few literatures explore room-scale indoor channels in the sub-band at 200~GHz. Furthermore, the aforementioned room-scale measurement campaigns consist of very few Tx-Rx positions, generally less than 20, due to the large time consumption with narrow beam scanning in the spatial domain. Therefore, an extensive room-scale channel study that focuses on the frequency band above 200~GHz and includes different indoor scenarios with abundant Tx/Rx-pair positions is still missing.

\par In this paper, we first present channel measurement campaigns conducted in a meeting room and an office room at 201-209~GHz, through a frequency-domain channel sounding method via a VNA. In particular, the two campaigns include four scenarios, namely, the meeting
room, the LoS cubicle area, the LoS hallway and the NLoS case. In light of the measurement results comprised of 90 Tx-Rx pairs, we study the large-scale path losses and small-scale channel parameters at 201-209~GHz in different indoor scenarios. Furthermore, we develop a ray-tracing-statistical hybrid model for the indoor environment, with proper parameter settings.
Our preliminary and shorter version in~\cite{he2021channel} reported path loss characteristics, and the developed single-frequency and multi-frequency path loss models in indoor scenarios at 201-209~GHz, in contrast with the results at 140~GHz. By making efforts on more extensive measurement results, study of power-delay-angular profiles and clustering results, and development of channel models, we summarize the contributions of this work as follows.
\begin{itemize}
    \item Extensive room-scale channel measurement campaigns are conducted in a meeting room and an office room. In total, 90 Tx-Rx pairs are measured with Rx scanning both azimuth and elevation angles, in the four distinctive scenarios, e.g., the LoS meeting room, the LoS cubicle area, the LoS hallway and the NLoS case. 
    \item We construct power-delay-angular profiles to investigate THz wave propagation features in different indoor scenarios. Specifically, temporal and spatial consistency in hallway and NLoS case are verified, offering two lessons. First, wireless propagation although at the same frequency, especially NLoS paths, exhibit noticeable differences in various indoor scenarios. Second, in light of the temporal and spatial consistency, beam tracking for THz communications is feasible.
    \item We characterize THz indoor channel thoroughly. On one hand, large-scale fading, i.e. best-direction and omni-directional path losses, in the four communication scenarios are compared and modeled based on close-in models. On the other hand, small-scale parameters, e.g., the number of clusters, delay spread, angular spread, and cluster excess delay are analyzed and statistically modeled. In addition, statistics of reflection loss and the relation between the NLoS cluster power and excess delay are characterized.
    \item We propose a general ray-tracing-statistical hybrid model framework for THz indoor channels. The deterministic part of the channel model captures the wall-reflection paths with the statistically modeled reflection loss. The statistical distributions and proper parameters for the meeting room, cubicle area, hallway and NLoS case at 200 GHz are extracted. The hybrid channel model is validated by the measured delay spread and angular spread in the four scenarios.
\end{itemize}
\par The remainder of this paper is organized as follows. In Sec.~\ref{Sec:ChannelMeasurementCampaign}, we describe the details of the THz channel measurement platform as well as the channel measurement campaigns in the four communication scenarios. Temporal and spatial consistency among different Rx positions in hallway and NLoS cases are verified and studied in Sec.~\ref{Sec:MPCsandSpatialConsistency}. Then, the large-scale propagation loss models are derived for path loss and reflection loss in Sec.~\ref{Sec:PathLossCharacterizationsandModels}. The small-scale fading channel parameters including the number of clusters, delay spread, angular spread, cluster time of arrival (ToA) in different indoor communication scenarios at 201-209~GHz are analyzed and modeled stochastically in Sec.~\ref{Sec:ChannelParameterCharacterizations}. A ray-tracing-statistical hybrid model is proposed and validated in Sec.~\ref{Sec:ChannelModel}. Finally, the paper is concluded in Sec.~\ref{Sec:Conclusion}.

\section{Channel Measurement Campaign} \label{Sec:ChannelMeasurementCampaign}
\begin{figure}[thbp]
\centering
    \subfigure[Meeting room]{\includegraphics[width=0.45\textwidth]{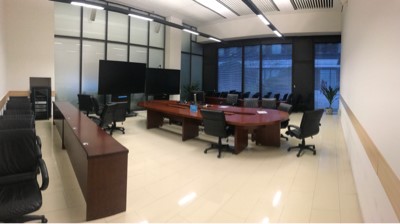}}
	\subfigure[Office room]{\includegraphics[width=0.45\textwidth]{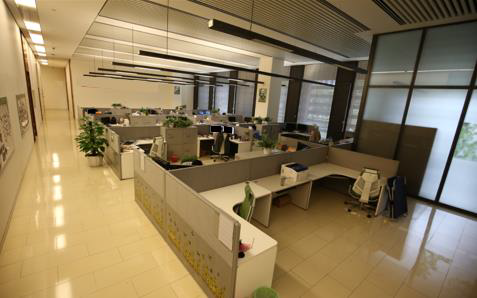}
	}
\caption{Photos of the (a) meeting room and (b) office room.}
\label{fig:photo}
\end{figure}

In this section, we describe the THz measurement campaign, including the specification of the hardware system, indoor environments as shown in Fig.~\ref{fig:photo}, and measurement deployment. Moreover, system calibration is carried out for eliminating the impact of the measurement system on the channel characterization.

\begin{table}
  \centering
  \caption{Parameters of the Measurement System.}
    \begin{tabular}{lll}
        \toprule
    Parameter & \multicolumn{1}{l}{Symbol} & Value \\
    \midrule
    Start frequency &   $f_{start}$    & 201~GHz \\
    End frequency &   $f_{end}$    & 209~GHz \\

    Bandwidth &   $B_w$   & 8~GHz \\
    Sweeping points &   $ N$  & 801\\
    Sweeping interval &    $\Delta f$   & 10~MHz \\
    Average noise floor &  $P_N$     & -140/-160~dBm \\
    Test signal power &   $P_{in}$    & 1~mW \\
    HPBW of transmitter & $HPBW^{Tx}$     & $60^\circ$ \\
    HPBW of receiver &  $HPBW^{Rx}$  & $10^\circ$ \\
    Antenna gain at Tx &   $G_{\text{t}}$   & 8 dBi \\
    Antenna gain at Rx &   $G_{\text{r}}$   & 25 dBi \\
    Time domain resolution &  $\Delta t$     & 125~ps \\
    Path length resolution &   $\Delta L$    & 3.75~cm \\
    Maximum excess delay &    $\tau_m$   & 100~ns \\
    Maximum path length &   $L_m$    & 30~m \\
    Azimuth rotation range &      & $[0^\circ:10^\circ:360^\circ]$ \\
    Elevation rotation range &    & $[-20^\circ:10^\circ:20^\circ]$ \\
        \bottomrule
    \end{tabular}%
  \label{tab:mparameters}%
\end{table}%

\begin{figure*}[thbp]
\centering
    \subfigure[Meeting room]{\includegraphics[width=0.5\textwidth]{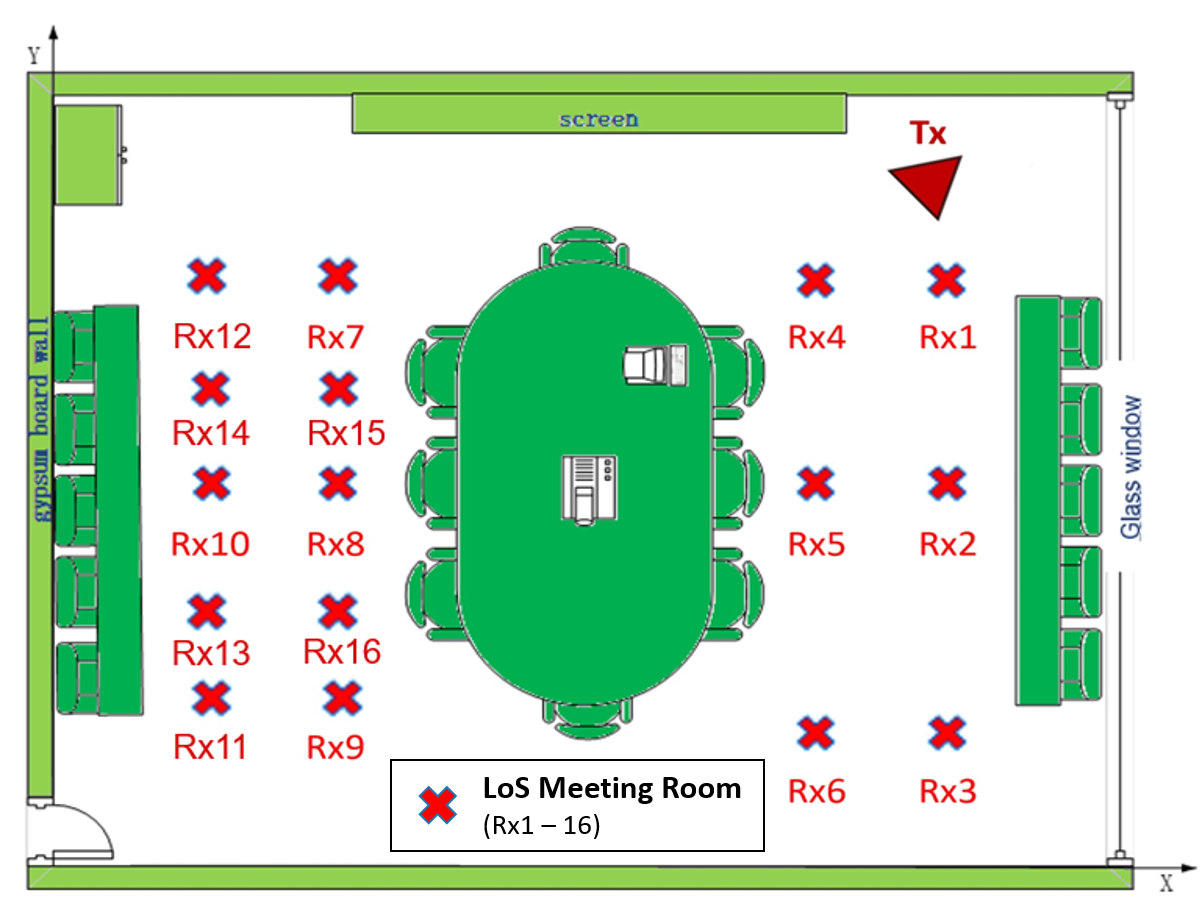}}
    \label{fig:deployment_meeting}
    
	\subfigure[Office room]{\includegraphics[width=0.9\textwidth]{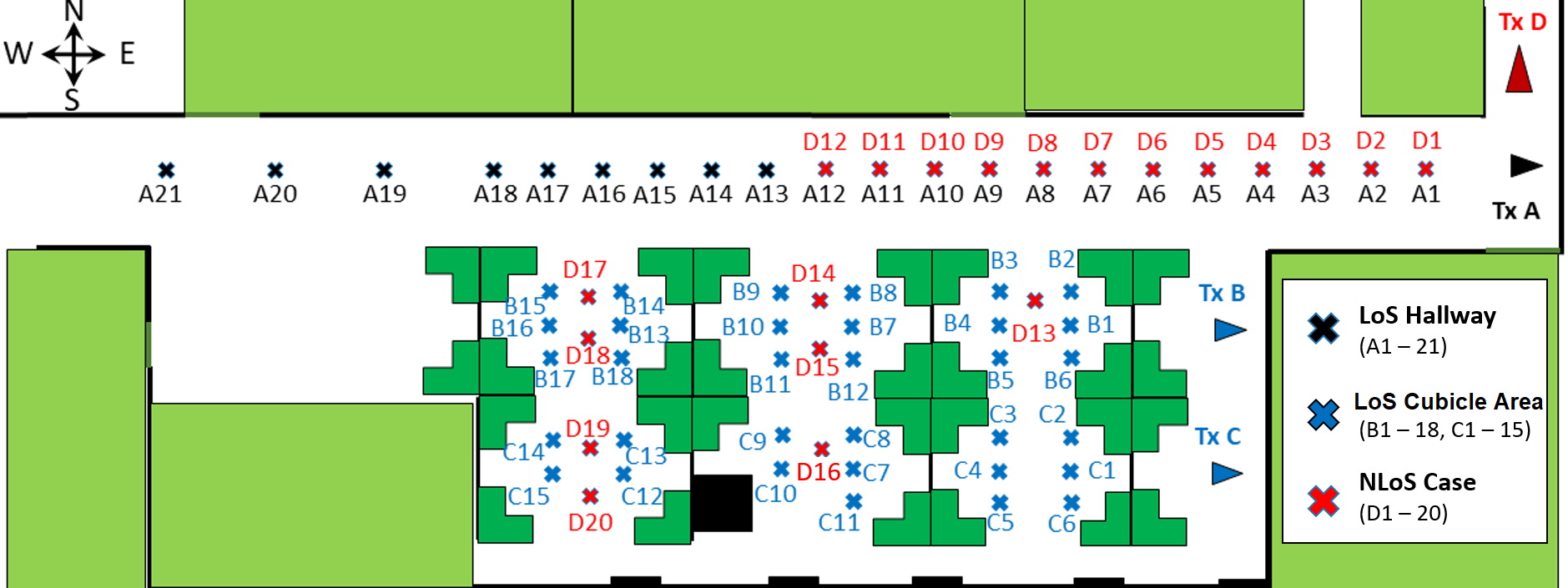}
	\label{fig:deployment_office}
	}
\caption{The deployment of the extensive channel measurement in (a) a meeting room and (b) an office room.}
\label{fig:deployment}
\end{figure*}

\subsection{Channel Measurement System at 201-209~GHz}
\par The THz channel measurement platform at 201-209~GHz consists of radio frequency (RF) fronts with horn antennas at both Tx and Rx sides supported by a VNA. The local oscillator (LO) signal of 18~GHz is multiplied by a factor of 12 to 216~GHz. The intermediate frequency (IF) signals generated by VNA range from 7~GHz to 15~GHz, which are mixed with the multiplied LO signal to the frequency band from 201 to 209~GHz. The measured bandwidth $B_w$ is 8~GHz. Therefore, the time domain resolution of our measurement results is 125~ps. A directional horn antenna at Tx produces the half-power beamwidth (HPBW) of $60^\circ$ with an antenna gain of around 8~dBi at 201-209~GHz, to guarantee a wide angular coverage. The Rx antenna gain is 25~dBi, and the HPBW is $10^\circ$, which is one-sixth of that at Tx for high spatial resolution. The Rx is mounted on a rotation unit, which can be rotated by step motors. In addition, the power of the test signal is 1~mW, while the noise floor in the temporal domain of our THz measurement platform is $-140$~dBm in the office room and $-160$~dBm in the meeting room, respectively. Detailed parameters of the measurement system are summarized in Table~\ref{tab:mparameters}. 


\subsection{Meeting Room Environment and Measurement Deployment}
We carry out the channel measurement in a typical meeting room with an area of 10.15~m~$\times$~7.9~m and a ceiling height of 5.8~m. In the meeting room, a 4.8~m~$\times$~1.9~m oval table with a height of 0.77~m is placed in the center, while eight chairs are around the oval table, as shown in Fig.~\ref{fig:deployment}(a). In addition, two televisions are closely placed in front of the north wall. The material of the east wall is glass, while the other three are made of lime. We notice that the maximum detectable path length imposed by the measurement system is 30~m, which is three times the dimension of the meeting room. As a result, reflected paths with at most third-order reflection can be detected in our measurement. 
\par  In our measurement deployment, 16 positions of Rx are placed in the meeting room, as depicted in the top view of the meeting room in Fig.~\ref{fig:deployment}(a). Tx is close to a corner of the meeting room while the Rx is placed on the positions of from Rx1 to Rx16. For the measurement of each Rx, the main beam of Tx is directed to the Rx position. By contrast, Rx with the spatial resolution of $10^\circ$ scans the receiving beam in the azimuth domain from $0^\circ$ to $360^\circ$ and the elevation domain from -$20^\circ$ to $20^\circ$ to detect sufficient multi-paths. Therefore, the considered reflected paths collected in our experiment are mainly from the oval table, chairs, and walls, whose elevation angles are sufficiently confined within [-$20^\circ$, $20^\circ$].

\subsection{Office Room Environment and Measurement Deployment}
The dimensions of the office room in our channel measurement campaign are 30~m~$\times$~20~m, as shown in Fig.~\ref{fig:deployment}(b), including a hallway and a cubicle area. In the north of the office room, there is a 30-meter-long hallway. In the cubicle area, the space is partitioned by plastic boards into individual personal zones. On each desk, there are two monitors as well as other work-related items. 

\par The measurement campaign consists of three sets, (i) LoS cubicle area, (ii) LoS hallway, and (iii) NLoS case. First, in the measurement set of LoS cubicle area, Tx is placed at Tx~B and Tx~C, respectively. When Tx is placed at Tx~B, Rx is placed at B1-B18. When Tx is placed at Tx~C, Rx is placed at C1-C15. Here, there are in total 33 measurement points in the LoS cubicle area, in which the distance between Tx and Rx varies from 3.5~m to 14~m. Second, in the measurement set of the LoS hallway, Tx is placed at Tx~A while Rx is placed at A1-A21. The distance between Tx and Rx ranges from 2~m to 30~m. Third, in the measurement set of the NLoS case, Tx is placed at Tx~D, which is behind the corner of the hallway. 20 measured Rx points locate at D1-D20 without the existence of an LoS path. The distance between Tx and Rx is 3.75~m-20~m. In the aforementioned measurement campaign inside the office, there are in total 74 measurement points. For each measurement point, the main lobe of Tx directs to the Rx in the LoS cases. In the NLoS case, Tx always directs to position A1. 

\subsection{Normalization and System Calibration}
Before conducting channel measurements, a normalization procedure is carried out to de-embed the frequency response of the cables. After channel measurements, system calibration needs to be conducted to eliminate the effect of the VNA, and RF fronts at Tx and Rx. The process of system calibration requires to first measure the channel transfer function (CTF) of a back-to-back connection with an attenuator, which is denoted by $S_{\text{calibration}}$, as
\begin{equation}
    S_{\text{calibration}}=H_{\text{attenuator}}H_{\text{system}},
\end{equation}
where $H_{\text{attenuator}}$ is the frequency response of the attenuator and $H_{\text{system}}$ is the frequency response of the measurement system. The measured S21 parameter from our channel measurement campaigns is, 
\begin{equation}
    S_{\text{measured}}=H_{\text{system}}H_{\text{channel}},
\end{equation}
where $H_{\text{channel}}$ is the realistic channel transfer function of THz signals in indoor scenarios. Therefore, by elimnating the effect of the system, The realistic channel transfer function of THz signals in indoor scenarios is represented as, 
\begin{equation}
    H_{\text{channel}}=\frac{S_{\text{measured}}H_{\text{attenuator}}}{S_{\text{calibration}}}.
\end{equation}

\begin{figure}
\centering
\includegraphics[width=0.6\textwidth]{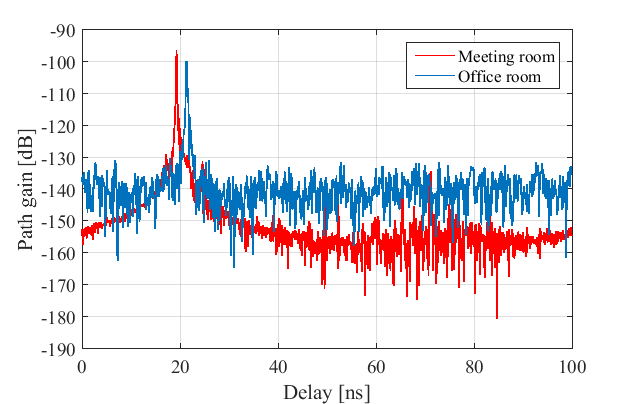}
\caption{Measured CIRs in the meeting room and office room at 201-209~GHz.}
\label{fig:CIR}
\end{figure}

\subsection{MPC Extraction and Clustering}
With 8~GHz bandwidth and a sweeping interval of 10~MHz, we obtain 801 samples of channel impulse response (CIR) after inverse discrete Fourier transform (IDFT) on the measured channel transfer function for one direction. As a result, we regard there are 801 MPCs in the temporal domain for a rotated direction. The reasons why we do not use Space-Alternating Generalized Expectation-maximization (SAGE) algorithm to estimate MPCs are twofold. First, the SAGE algorithm requires a very accurate 3D antenna pattern, which is hard to obtain. Second, the SAGE algorithm suffers from phase inaccuracy due to cable swing during channel measurement campaigns. For each measured Tx-Rx pair, the threshold for noise elimination is given as
\begin{equation}
    \text{TH} \text{ [dB]}=\max(P_m-40,\text{NF}+10),
\end{equation}
where $P_m$ is the largest power among measured MPCs and $\text{NF}$ is the estimated noise floor in dB.
\par The signals are transmitted through coaxial cables, where the transmission loss of IF signals increases with frequency and transmission distance. Therefore, the frequency response of cables is normalized by automatically compensating power at higher frequencies in the normalization procedure. The noise floor is thereby lifted up at higher frequencies. Moreover, the measurement in the office requires longer coaxial cables. As a result, the noise floor of measured channels in the office room is higher than that in the meeting room. Fig.~\ref{fig:CIR} shows the channel impulse responses in the meeting room and the office room, respectively. We observe that the noise floor in the office room (-140~dB) is about 20~dB higher than that in the meeting room (-160~dB).
\par The clustering algorithm we adopt is Density-Based
Spatial Clustering of Applications with Noise (DBSCAN)~\cite{chen2021channel}. The advantages of DBSCAN includes automatic determination of cluster numbers, finding arbitrary cluster shape, and robustness to the outliers. Multipath component distance (MCD) is used in the DBSCAN algorithm to measure the distance between two MPCs. The two required parameters of DBSCAN, i.e., the minimum number of points in a cluster and the minimum distance between two clusters, are empirically set to be 5 and 0.05, respectively.

\section{Multipath Propagation and Spatial Consistency} \label{Sec:MPCsandSpatialConsistency}
In this section, we analyze the multipath propagation in the indoor scenarios via the measured power-delay-angular profiles (PDAPs) in different scenarios at 201-209~GHz. Moreover, temporal and spatial consistency among different Rx positions in the LoS hallway and the NLoS cases are elaborated, which supports the beam-tracking technique for indoor THz communication.

\begin{figure}
\centering
\includegraphics[width=0.6\textwidth]{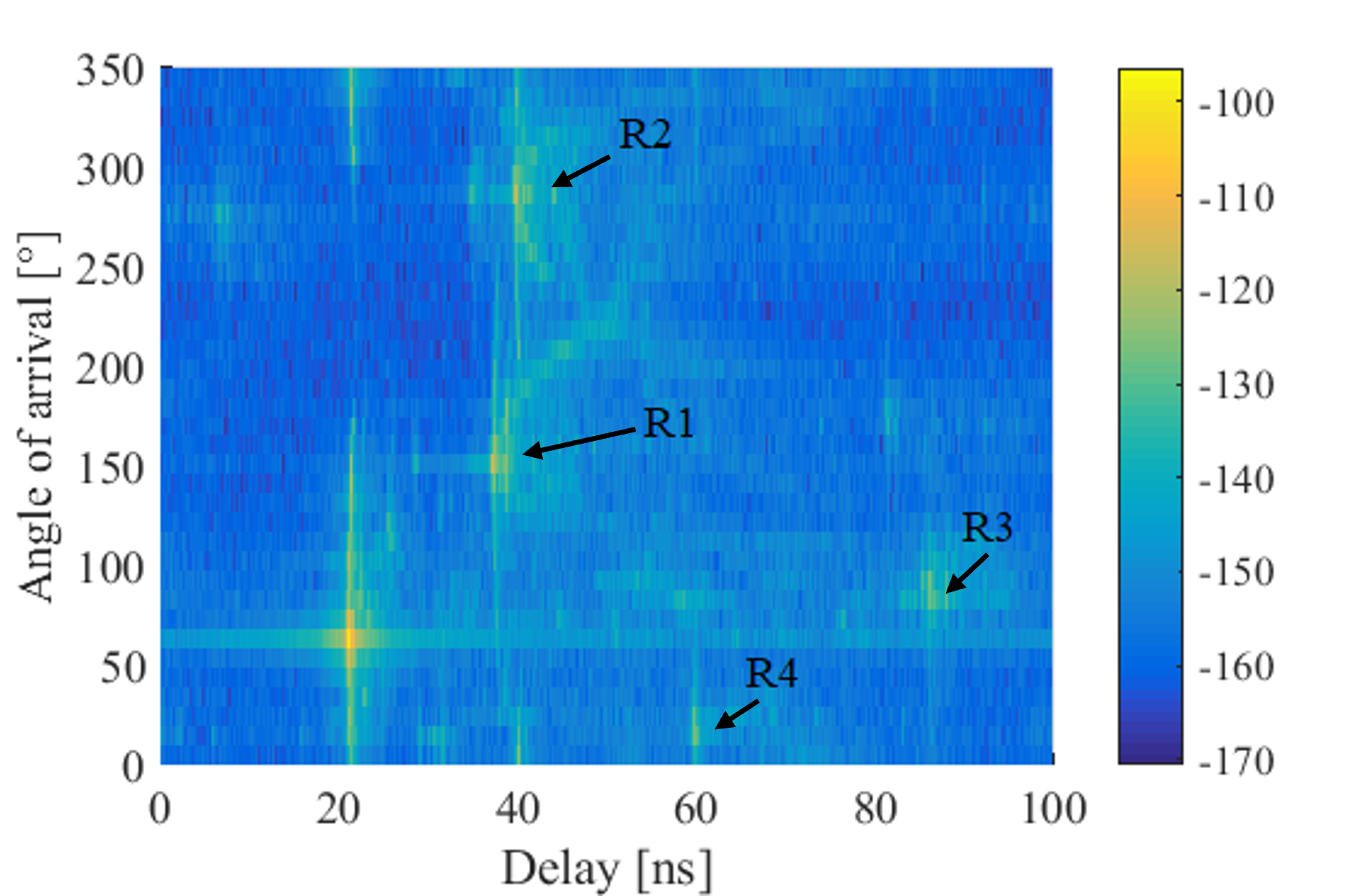}
\caption{Measured PDAP at Rx8 in the meeting room at 201-209~GHz.}
\label{fig:meeting_rx8}
\end{figure}

\subsection{LoS Meeting Room}
Fig.~\ref{fig:meeting_rx8} shows the PDAP at Rx8 in the meeting room, where different significant propagation paths are described in black. In particular, the significant MPCs include LoS path and paths reflected from the walls, which are consistent with our observations on the channel measurement campaign at 140~GHz~\cite{yi2021Channel}. To be concrete, R1 and R2 correspond to the first-order reflection on the south wall and west wall, respectively. The received power of those two paths is stronger than others, as they are within the mainlobe of Tx as well as have shorter propagation distances. R3 relates to the second-order reflections on the west wall and the east wall. R4 is the path that first reflects on the south wall and then on the west wall before arriving at Rx. R3 and R4 are relatively weak due to higher reflection orders. As a result, we state that characterizing the wall-reflected paths is very critical in the channel modeling in the meeting room in the THz band.

\begin{figure}
\centering
\includegraphics[width=0.6\textwidth]{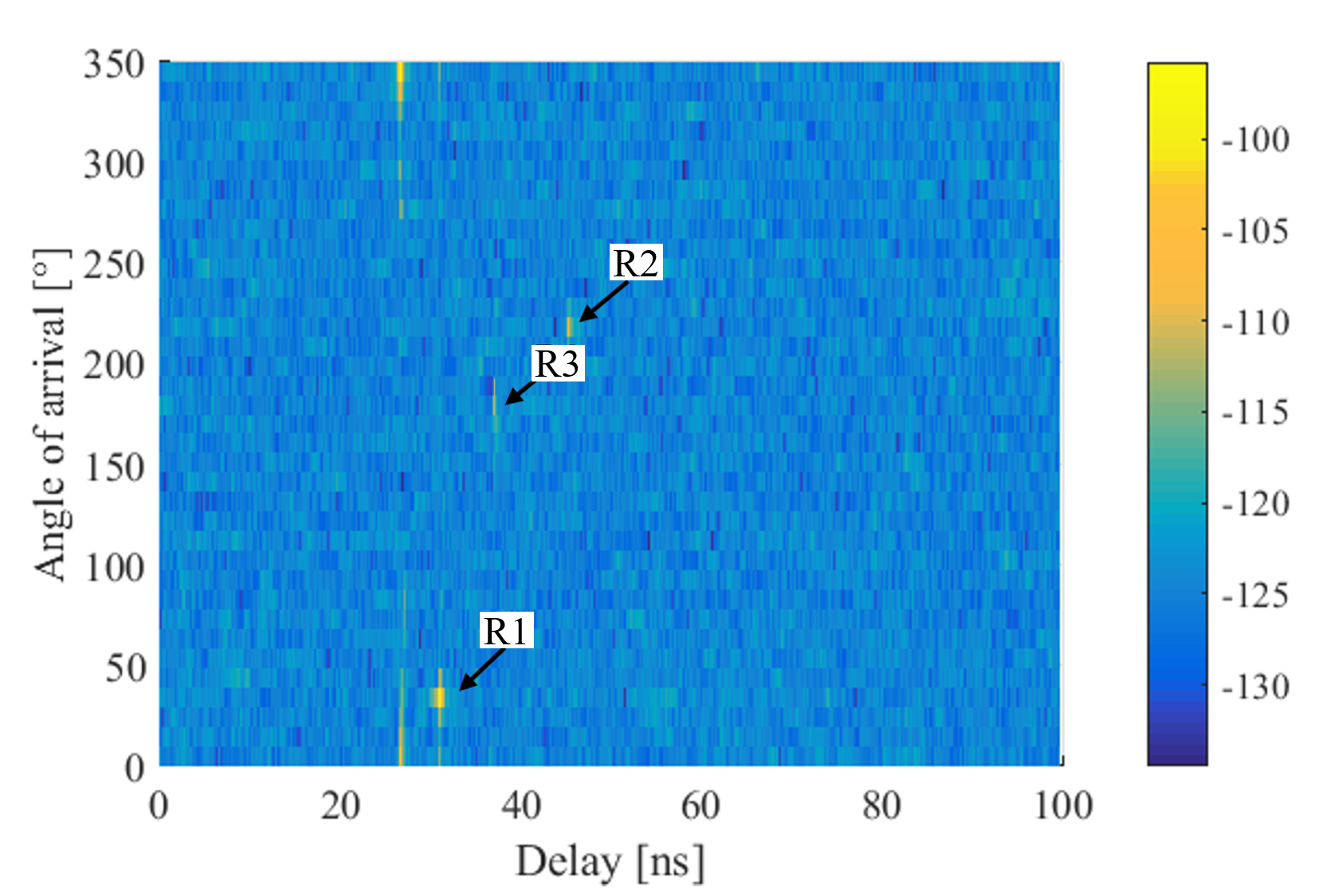}
\caption{Measured PDAP at C11 in cubicle area at 201-209~GHz.}
\label{fig:meeting_rx11}
\end{figure}

\subsection{LoS Cubicle Area}
Different from the propagation in the meeting room, significant paths in the cubicle area are resulted from reflection on the partitions or monitors. We take C11 as an example, whose PDAP is plotted in Fig.~\ref{fig:meeting_rx11}. As described in black, R1 propagates from the east-south partition while R2 reflects from the west-north partition. In addition, R3 reflects back from the pillar. The reasons that wall-reflection paths are missing in the cubicle area are twofold. First, the dimension of the office room is much larger than the meeting room, which causes the mainlobe of Tx cannot fully cover the wall-reflection paths. Second, Rx in the cubicle area is surrounded by the partitions and scatterers, which produce reflections. As a result, we observe that although the meeting room and cubicle area are both indoor scenarios, multipath propagation in these two scenarios is significantly different.

\begin{figure}
\centering
\includegraphics[width=0.6\textwidth]{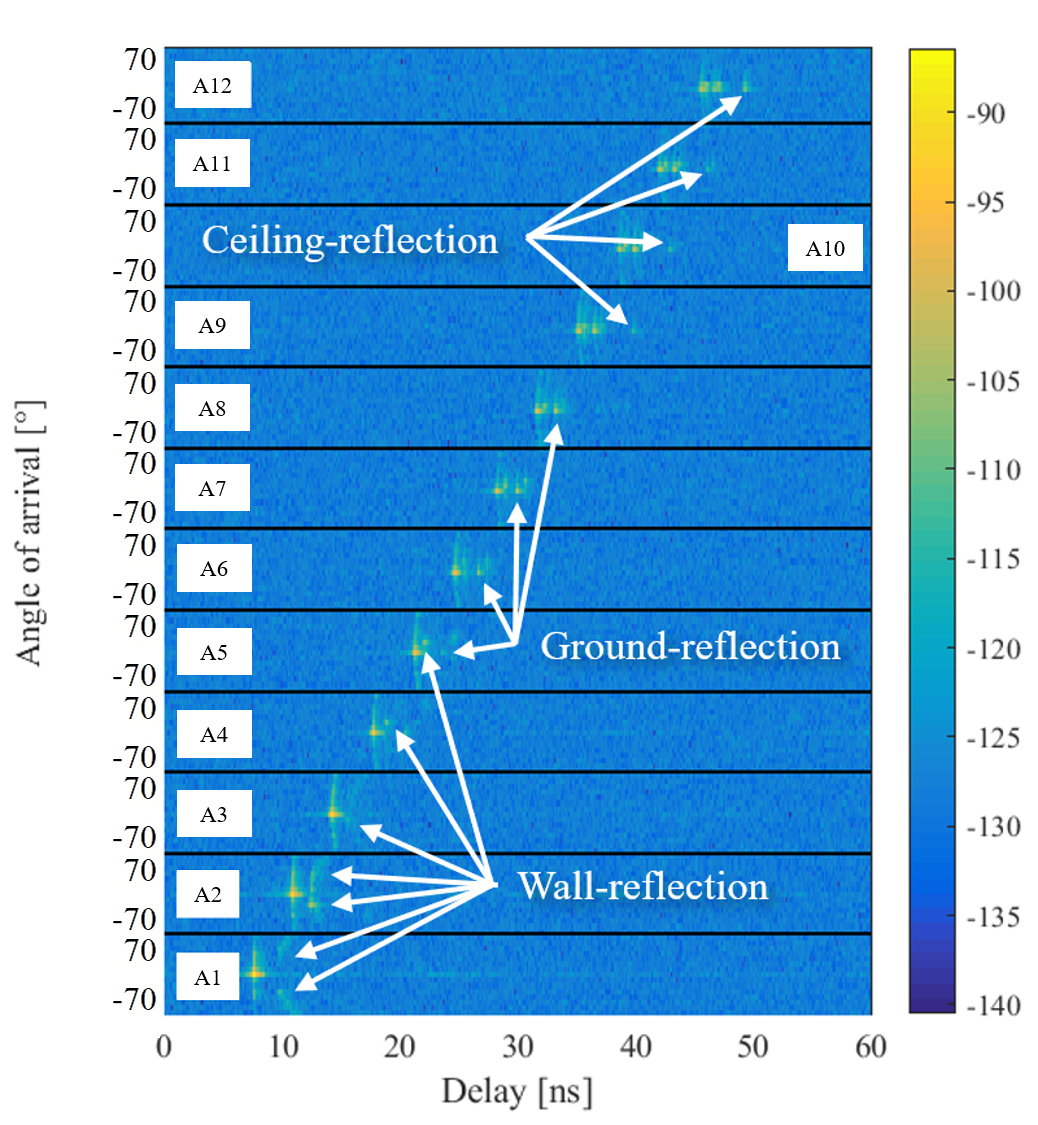}
\caption{Temporal and spatial consistency in hallway at 201-209~GHz.}
\label{fig:sc_hallway}
\end{figure}

\subsection{LoS Hallway}\label{sec:MPC_LoSHallway}
Twelve PDAPs of A1 to A12 in the LoS hallway scenario with azimuth AoAs from $-70^\circ$ to $70^\circ$ and ToAs from 0 to 60~ns are grouped and depicted in Fig.~\ref{fig:sc_hallway}. As Tx-Rx distance increases with a fixed increment from A1 to A12, the delays of the LoS path along with those reflection paths are observed to linearly increase (aligned in a line in the measured PDAPs). In each PDAP, one LoS path and some reflection paths can be observed. The reflection paths propagate from the wall, ground, and ceiling with several orders of reflection. Their AOA and delay are very similar to the LoS path, due to the waveguide effect in the hallway scenario. Moreover, as the Tx-Rx distance increases, the separation between the reflection paths and the LoS path in the PDAP shrinks. To explain this observation, we denote $L_{\text{LoS}}$ as the propagation length of the LoS path and $L_{r}$ as the distance between the Tx and the wall, ground, or ceiling. Then, the difference of the propagation distance between the LoS path and the reflection path is given by,
\begin{equation}
\begin{split}
    \Delta L&=\sqrt{L^2_{\text{LoS}}+(2L_{r})^2}-L_{\text{LoS}}\\
    &=\frac{4L_r^2}{\sqrt{L_{\text{LoS}}^2+4L_r^2}+L_{\text{LoS}}},
\end{split}
\end{equation}
where we can find that $\Delta L$ decreases with $L_{\text{LoS}}$, and $\Delta L$ is small if $L_r$ is small, which are consistent with the observations in PDAPs. 
\par The occurrence order of the mentioned reflection paths (from A1 to A12) starts from the wall-reflection paths, ground-reflection paths, and finally the ceiling-reflection paths. For example, wall-reflection paths are observed starting from A1. The reason is that the elevation angle of departure angle of the wall-reflection path is the same as that of the LoS path, and it is always within the mainlobe of Tx. Beyond A5, the wall-reflection paths cannot be distinguished from the loS path since the distance between the Tx and the wall, $L_r$ is only 0.8~m, which is small compared with the LoS path, $L_{\text{LoS}}>3$~m. In addition, the difference of the azimuth AOA between the LoS path and the wall-reflection path is calculated as,
\begin{equation}
\begin{split}
    \Delta \phi&=\arctan{(\frac{2L_r}{L_{LoS}})},
\end{split}
\end{equation}
which is observed to decreases with $L_{\text{LoS}}$.
\par The ground-reflection path occurs starting from A5 as shown in Fig.~\ref{fig:sc_hallway}. The reason is that when Rx is close to Tx, the elevation angle of departure of this path is out of the mainlobe of Tx. Moreover, this path combines with the LoS path at A9 and thereafter. The last reflection path to appear is the ceiling-reflection path at A9, since the distance between Tx and the ceiling is much larger than the distance between Tx and the ground. When the Tx and Rx are separated far enough, the elevation angle of departure for the ceiling-reflection path appears within the mainlobe of the Tx antenna. 
\par It should be noted that there are other NLoS paths in the hallway scenario, apart from the LoS path and reflection paths referring to the wall, ground, and ceiling that we mentioned above. Those NLoS paths include one that reflects back from the glass door and others caused by the scatterers, including the partitions, desks, and chairs in the cubicle area. Generally, those paths are weak and their AoA and ToA show a great difference from the LoS path. We conclude that the LoS path dominates in the hallway scenario, and the wall, ground, and ceiling reflection paths are likely to be combined with the LoS path and become unnoticeable in the measured PDAPs.

\subsection{NLoS Case}
\begin{figure}
\centering
\includegraphics[width=0.6\textwidth]{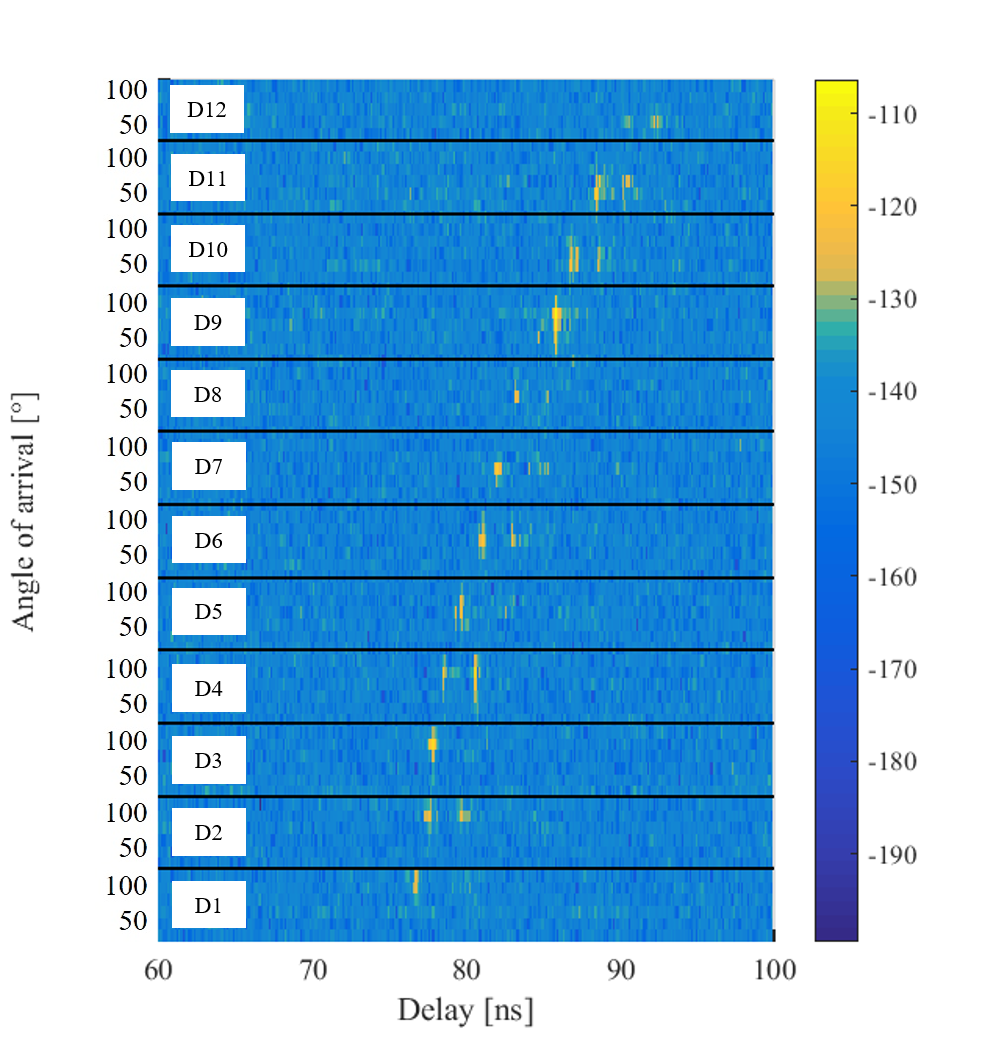}
\caption{Temporal and spatial consistency in NLoS case at 201-209~GHz.}
\label{fig:sc_nlos}
\end{figure}

Fig.~\ref{fig:sc_nlos} illustrates the PDAPs at D1 to D12 with ToA from 60 to 100~ns and azimuth AoA from $50^\circ$ to $100^\circ$ in the NLoS case. It should be clarified that Rx positions at D1 to D12 are the same as A1 to A12 in the LoS hallway scenario. In this case, the transmitter is placed at the corner behind the hallway, as depicted in Fig.~\ref{fig:deployment_office}, where the boresight of Tx and Rx are blocked. Some NLoS paths with temporal and spatial consistency when Rx moves from D1 to D12 can be observed in the measured PDAPs. Different from the LoS hallway scenario, NLoS paths are not aligned in a line in Fig.~\ref{fig:sc_nlos}. In most of the PDAPs, we can find a first-arrival NLoS path followed by another NLoS path. The propagation of the two NLoS paths is very similar and the only difference is that the later-arrival one travels through an additional reflection on the ceiling. Besides, the azimuth AoAs of the NLoS paths clearly decrease as Rx moves farther from Tx. As a result, we observe that even in the NLoS case, beam tracking is possible at 201-209~GHz in the office room due to the continuous occurrence of the NLoS paths~\cite{chen2021millidegree}.

\section{Large-scale Propagation Loss Models}  \label{Sec:PathLossCharacterizationsandModels}
In this section, we first introduce the single-frequency CI path loss model. Then, we calculate the path loss in different scenarios from the channel measurement campaigns, and the CI path loss models are developed. In particular, the best-direction and omni-directional path losses are considered. Finally, reflection loss in the indoor environment is studied. 

\begin{figure*}[htbp]
\centering
\subfigure[Best-direction path loss and single-frequency CI models.]{
\includegraphics[width=0.6\textwidth]{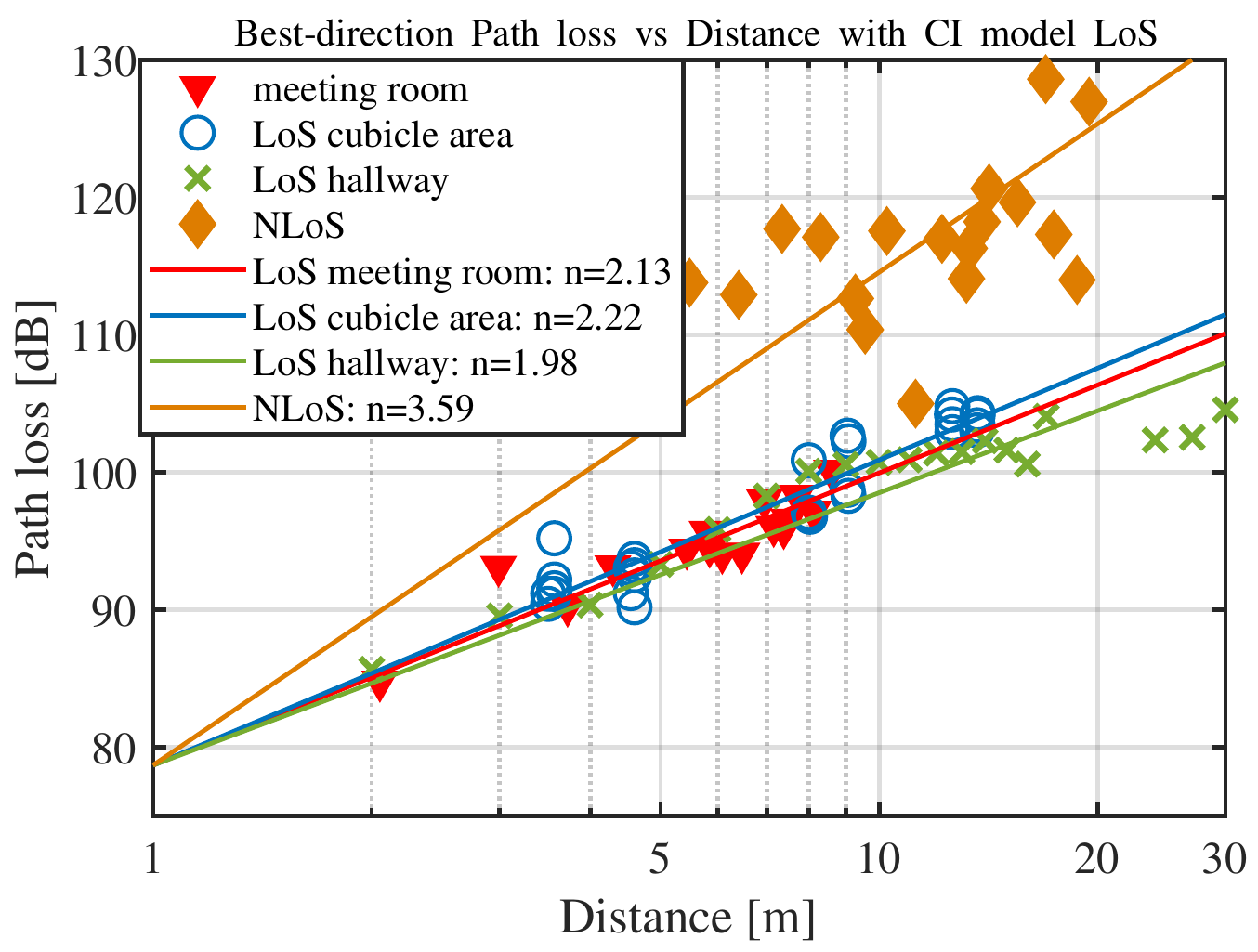}
    \label{fig:pl_best}
}
\subfigure[Omni-directional path loss and single-frequency CI models.]{
\includegraphics[width=0.6\textwidth]{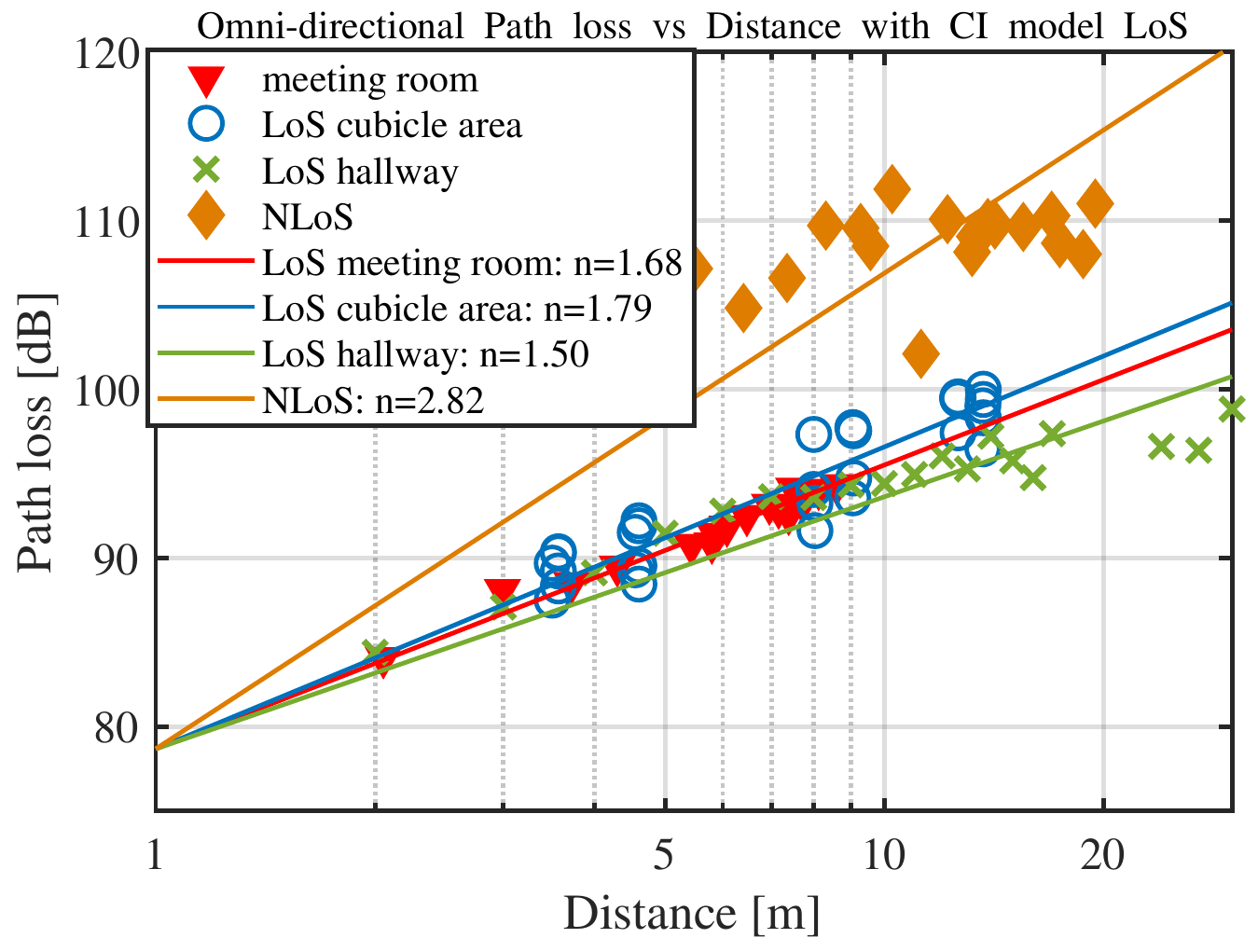}
\label{fig:pl_omni}
}
\caption{(a) Best-direction and (b) omni-directional path loss measurement results and single-frequency CI models in four scenarios at 201-209~GHz.}
\label{fig:pl} 
\end{figure*}

\subsection{Close-in (CI) Path Loss Model}
Path loss is a large-scale fading parameter which reveals the signal power attenuation at different places of Rx. We evaluate the CI path loss model for all the measurement sets, respectively. In general, the CI path loss model is represented as,
\begin{equation}
\text{PL}^{\text{CI}}[\text{dB}]=10\ \text{PLE}\ \log_{10}{(\frac{d}{d_0})}+\text{FSPL}(d_0)+X^{\text{CI}}_\sigma,
\label{eq:CI}
\end{equation}
where PLE is the path loss exponent, $d$ denotes the distance between Tx and Rx, and $d_0$ represents the reference distance which is 1~m in this work. $X^{\text{CI}}_\sigma$ is a zero-mean Gaussian random variable with standard deviation $\sigma^{\text{CI}}_{\text{SF}}$ in dB, which represents the fluctuation caused by shadow fading. Moreover, we compute the free-space path loss (FSPL) by invoking the Friis' law, given by
    \begin{equation}
    \text{FSPL}(d_0)=-20\log_{10}(\frac{c}{4\pi fd_0}),
    \label{eq:fspl}
    \end{equation}
where $c$ denotes the speed of light, and $f$ represents the carrier frequency. In addition, PLE in \eqref{eq:CI} is determined by minimizing $\sigma^{\text{CI}}_{\text{SF}}$ via a minimum mean square error (MMSE) approach.

\subsection{Best-direction Path Loss}
In directional antenna channel measurements, the path loss can be classified into best-direction and omni-directional types. First, we consider the best-direction path loss as the directional path loss. That is, for each Tx-Rx position, we only calculate the path loss received from the best direction which has the strongest received power. For the LoS case, it is calculated as,
\begin{equation}
    \text{PL}^{\text{LoS}}_{best}=-10*\log_{10}{(\max_{i,j}{|H_{i,j}|^2_{\text{avg}}})},
    \label{los-pl}
\end{equation}
where $|H_{i,j}|^2_{\text{avg}}$ is the average square of CTF over the measured frequency band at the $i^{\text{th}}$ azimuth angle and the $j^{\text{th}}$ elevation angle at Rx. The calculation of ${H^{\text{avg}}}_{i,j}$ is given by,
\begin{equation}
    {|H_{i,j}|^2_{\text{avg}}}=\sum_{s=1}^{S}\frac{|H_{i,j,s}|^2}{S},
\end{equation}
where $H_{i,j,s}$ is the CTF at the $s^{\text{th}}$ swept frequency at the $i^{\text{th}}$ azimuth angle and the $j^{\text{th}}$ elevation angle at Rx. $S$ denotes the number of swept frequencies.
\par By contrast, for the NLoS case, the path loss calculation is different from the LoS case in (9) and (10). The reason is that measured MPCs are weak and their power may be comparable to the noise floor in the NLoS case. The noise floor would dominate the calculated path loss given by \eqref{los-pl}, which is smaller than the realistic path loss. To eliminate the noise floor, we first transfer the CTF to obtain CIR, $h_{i,j,s}=\text{IDFT}(H_{i,j,s})$, and sort the CIR in a descending order of absolute values, $h^{\text{sorted}}_{i,j,s}$. As a result, the best-direction path loss in the NLoS case is then calculated as,
\begin{equation}
    \text{PL}^{\text{NLoS}}_{best}=-10*\log_{10}{(\max_{i,j}{P^{\text{AOA}}_{i,j}})},
\end{equation}
where $P_{i,j}$ denotes the power of the strongest $W$ MPCs at the $i^{\text{th}}$ azimuth angle and the $j^{\text{th}}$ elevation angle at Rx, as
\begin{equation}
    P^{\text{AOA}}_{i,j}=\sum_{s=1}^{W}|h_{i,j,s}^{\text{sorted}}|^2,
\end{equation}
where $W$ stands for the window size and is set to be 50 in our calculation.

\subsection{Omni-directional Path Loss}
\par Different from the best-direction path loss, the omni-directional path loss for each Tx-Rx position takes into account the power received from all the scanned angles at Rx, which is calculated as,
\begin{equation}
    \text{PL}_{omni}=-10*\log_{10}{(\sum_{i,j}{P^{\text{AOA}}_{i,j}})}.
\end{equation}
The omni-directional path loss is generally lower than the best-direction counterpart, as it involves all the received power.

\begin{table}
  \centering
  \caption{PLE of CI models for four indoor scenarios at 201-209~GHz.}
    \begin{tabular}{|c|c|c|c|c|}
    \hline
    Path loss type  & Meeting & \tabincell{c}{Cubicle\\area} & Hallway & NLoS \\
    \hline
    Best-direction & 2.13  & 2.22  & 1.98  & 3.59 \\
    \hline
     Omni-directional & 1.68  & 1.79  & 1.50  & 2.82 \\
    \hline
    \end{tabular}%
  \label{tab:ple_201-209}%
\end{table}%
The best-direction and omni-directional path loss for four indoor scenarios at 201-209~GHz are shown in Fig.~\ref{fig:pl}, while the corresponding PLE of CI models are summarized in Table~\ref{tab:ple_201-209}.
The observations and analysis are drawn as follows. First, since high-directional antennas are utilized at Rx, the best-direction propagation in the LoS hallway scenario is approximately equivalent to the free-space propagation, and the value of PLE in this case is 1.98, which is close to 2, as the PLE for free-space path loss.
Besides, the PLE value which is slightly lower than 2 can be explained by the waveguide effect.
Second, as expected for the NLoS scenario, the omni-directional PLE is about 0.5 higher than best-direction PLE in the other three scenarios, since the omni-directional path loss sums up the received power from all directions.
By contrast, in the NLoS case, noise power dominates the received power, i.e., received power from all directions is all comparable to the noise floor. Therefore, the omni-directional received power and thus the omni-directional PLE are much smaller than the best-direction counterparts.
Furthermore, the PLE in the cubicle area is slightly larger than the PLE in the meeting room, since strongly reflected paths from the walls are out of the main beam of Tx in the cubicle area, due to the larger dimensions of the office room than the meeting room~\cite{yi2021Channel}.
 \begin{figure}
\centering
\includegraphics[width=0.6\textwidth]{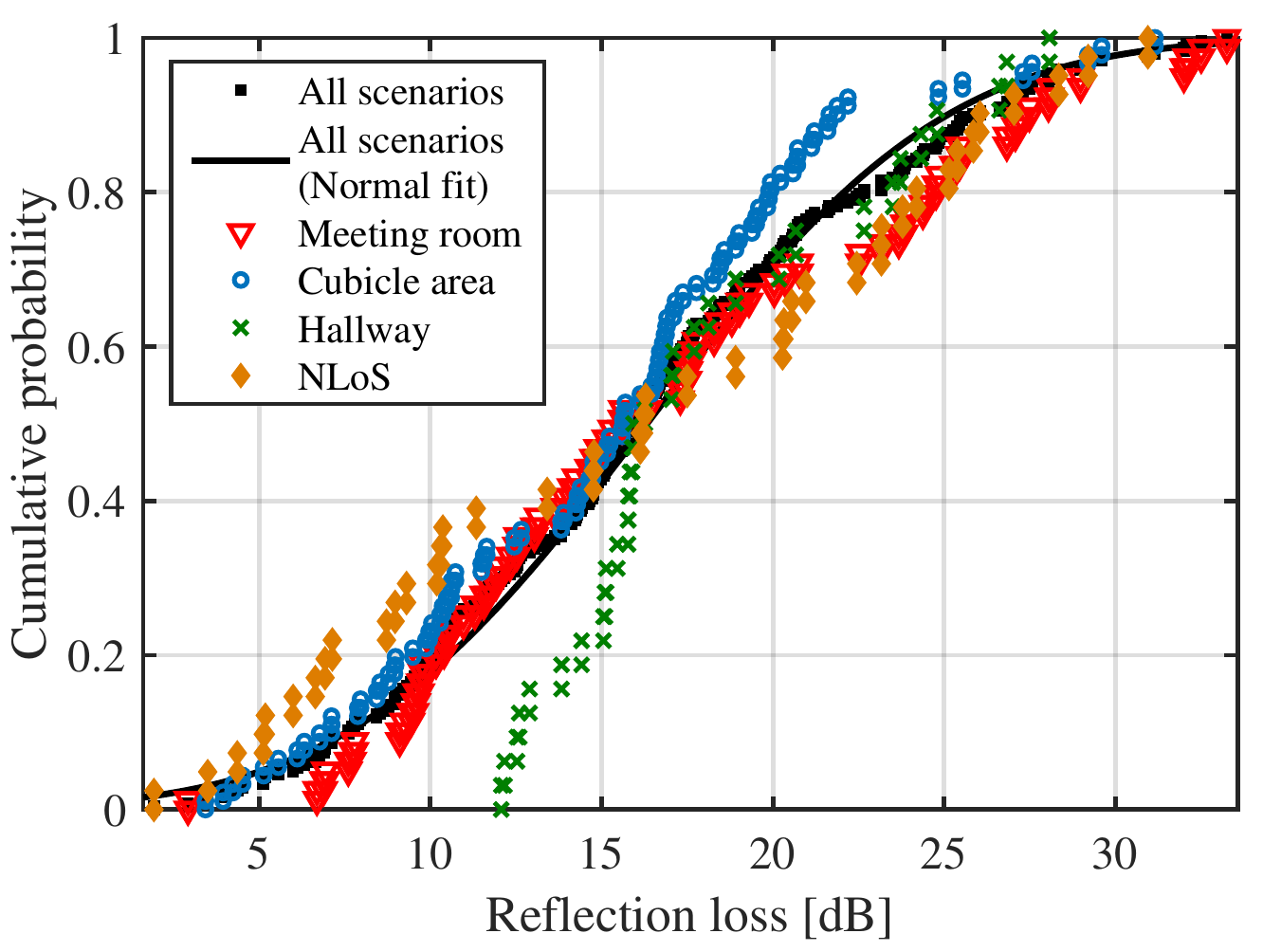}
\caption{Reflection loss in four scenarios and normal fitting at 201-209~GHz.}
\label{fig:refl_loss}
\end{figure}
\subsection{Reflection Loss}
\label{sec:reflection-loss}
Modeling reflection loss is critical to evaluate the received power of reflection paths in indoor scenarios at 201-209~GHz. In this work, we regard a cluster corresponding to a reflection path and define the reflection loss, $\text{RL}$, as,
\begin{equation}\label{eq:rl}
    \text{RL} \text{[dB]}=10\log_{10}(\frac{P_i^c}{ P_t})-10\log_{10}{(4\pi f\tau_i^c)},
\end{equation}
where $P_i^c$ is the received power of the $i^{\text{th}}$ cluster, $P_t$ denotes the transmit power, and $\tau^c_i$ represents the ToA of the $i^{\text{th}}$ cluster. $\tau_i^c $ is defined as the ToA of the MPC with the largest received power in the $i^{\text{th}}$ cluster. One should note that the loss due to the Tx antenna misalignment and multiple reflections are taken into account in the calculated reflection loss in \eqref{eq:rl}.
\par As shown in Fig.~\ref{fig:refl_loss}, the distributions of reflection loss in the indoor scenarios are very similar, since the materials of reflection surfaces in these scenarios are common. However, the lower bound of reflection loss in the hallway scenario is much higher than in the other three scenarios. The reason is that some strong reflection paths combine with the LoS path and can not be distinguished, as we explained in Sec.~\ref{sec:MPC_LoSHallway}, and the corresponding small reflection loss values are not taken into account. In addition, the reflection loss in the cubicle area is slightly lower than in the meeting room, since the metallic partitions and liquid-crystal monitor on the desks around the Rx produce strong reflections. This can explain that the office room experiences severer power dispersion in both delay and angular domains than the meeting room, as we observed in Sec.~\ref{sec:channel_ds} and Sec.~\ref{sec:channel_asa}. We also plot the reflection loss over all the scenarios in black cubes in Fig.~\ref{fig:refl_loss}. The reflection loss over all scenarios is well fitted by a log-normal distribution with $\mu_{\ln(\text{RL})}=2.71$ (i.e., 14.88~dB) and $\sigma_{\ln(\text{RL})}=0.50$.

\section{Small-scale Channel Parameters Characterization} \label{Sec:ChannelParameterCharacterizations}
In this section, we analyze the critical small-scale channel parameters, including the statistics of the number of clusters, delay spread, angular spread, as well as the relation between the cluster power and excess delay.
\begin{figure}
\centering
\includegraphics[width=0.6\textwidth]{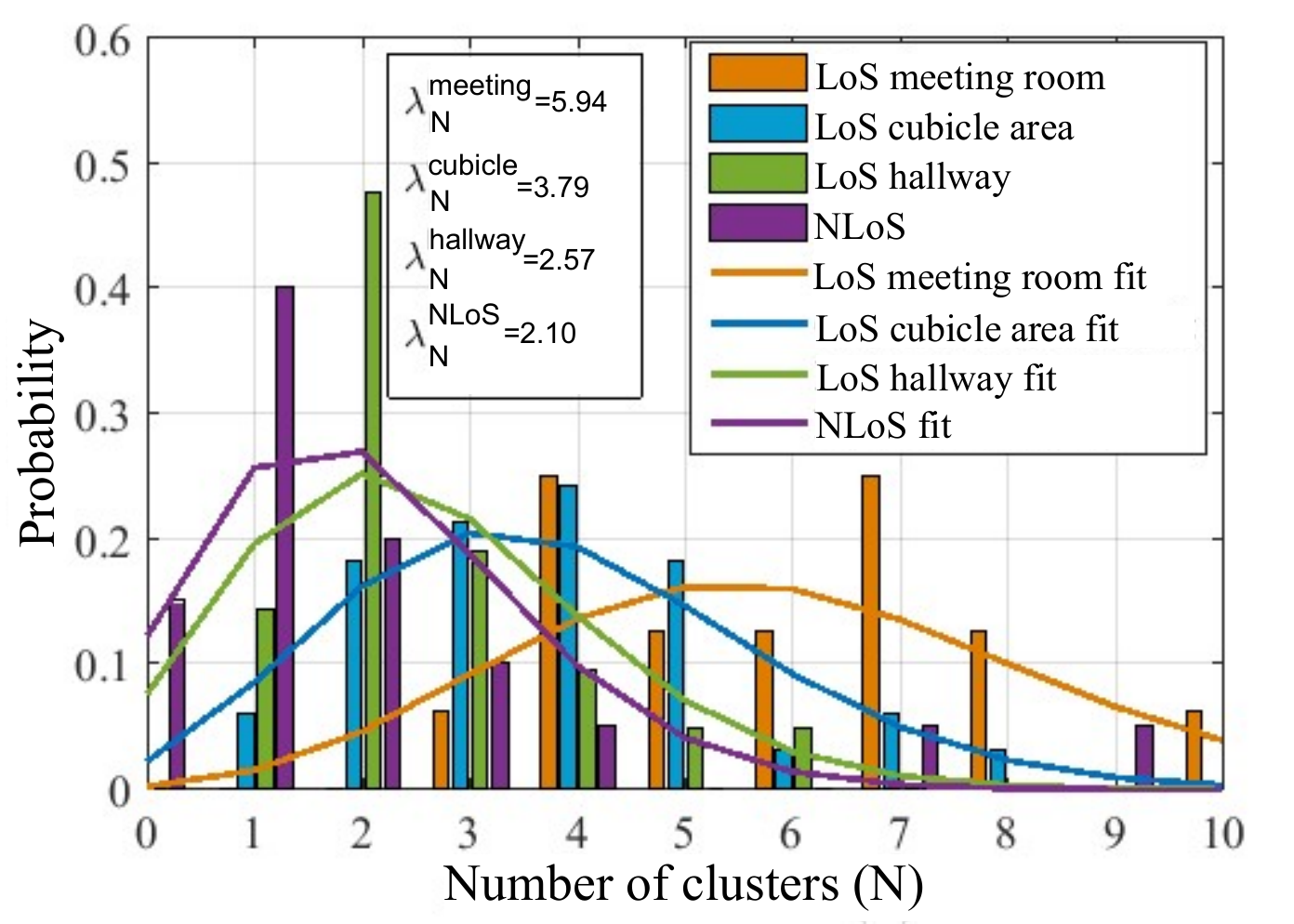}
\caption{Number of clusters and Poisson fitting in four scenarios at 201-209~GHz.}
\label{fig:N}
\end{figure}

\subsection{Number of Clusters}
The probability of the number of clusters for various indoor scenarios at 201-209~GHz is depicted in Fig.~\ref{fig:N}. 
First, the number of clusters is typically smaller than 10 in all indoor scenarios, indicating the sparsity of typical indoor channels.
Besides, the NLoS scenario reveals the smallest average number of clusters in the given interval due to the blockage of boresight between Tx and Rx. Except for the NLoS case, all the LoS scenarios witness at least one cluster. Especially, the minimum number of clusters in the meeting room is 3, due to the rich scattering environment, the contribution of wall-reflection paths, and the lower noise floor.
Moreover, the number of clusters in the cubicle area is smaller than that in the meeting room, since the office has a larger dimension and the distance between Tx and Rx is larger than that in the meeting room.
\par The number of clusters in each scenario is fitted by a Poisson distribution, as shown in Fig.~\ref{fig:N}. The average values of the number of clusters, $\lambda_N$, are calculated as 5.94, 3.79, 2.57 and 2.10 for the meeting room, cubicle area, hallway, and NLoS case, respectively. The hallway scenario shows a fewer number of clusters, due to the fact that some of the reflection paths are combined with the LoS cluster. The average values are smaller than 6, which is the suggested number of clusters given in 3GPP TR38.901 for indoor scenarios and frequencies below 100~GHz~\cite{3gpp.38.901}. This indicates that the low-Terahertz channel is sparser than the mmWave channel. We emphasize that the number of clusters in our measurements would increase if an omni-directional antenna is used at Tx.

\begin{figure}
\centering
\includegraphics[width=0.6\textwidth]{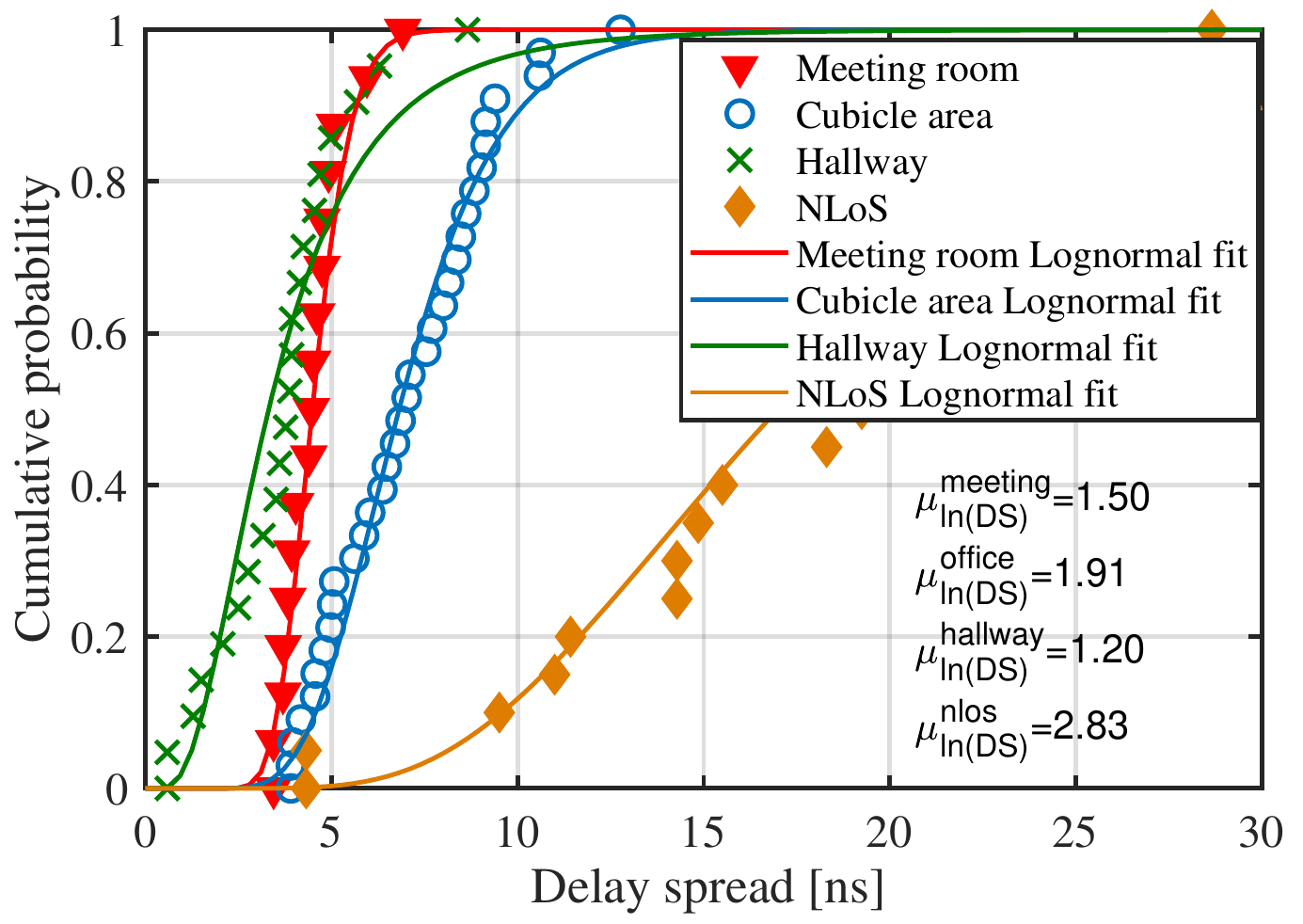}
\caption{Delay spread distribution of the measured channel in four scenarios at 201-209~GHz.}
\label{fig:ds_201-209}
\end{figure}

\subsection{Delay Spread}\label{sec:channel_ds}
Delay spread characterizes the power dispersion of MPCs in the delay domain. Root-mean-square (RMS) DS in different scenarios at 201-209~GHz and the log-normal distribution fitting are presented in Fig.~\ref{fig:ds_201-209}. The average log values of measured RMS DS in the meeting room, cubicle area, hallway, and NLoS case are, 1.5, 1.91, 1.2, and 2.83, respectively.
The hallway scenario shows the lowest RMS DS among the four scenarios, since the ToAs of the strong reflection paths are similar to the LoS path.
Moreover, the NLoS case derives the highest RMS DS. The reason is that the LoS path with strong received power is blocked, while the remaining paths are NLoS paths with weak received power and a large difference in ToA.
One interesting observation is that although the average value of the number of clusters in the meeting room is larger than that in the cubicle area, the average RMS DS in the meeting room is by contrast smaller than that in the cubicle area. This is due to the fact that the noise floor in the meeting room is lower than that in the office room and thus more reflection paths can be included in the meeting room, which results in a larger number of clusters. However, NLoS paths in the meeting room are much weaker compared with the LoS path. which causes the smaller RMS DS. 

\begin{figure}
\centering
\includegraphics[width=0.6\textwidth]{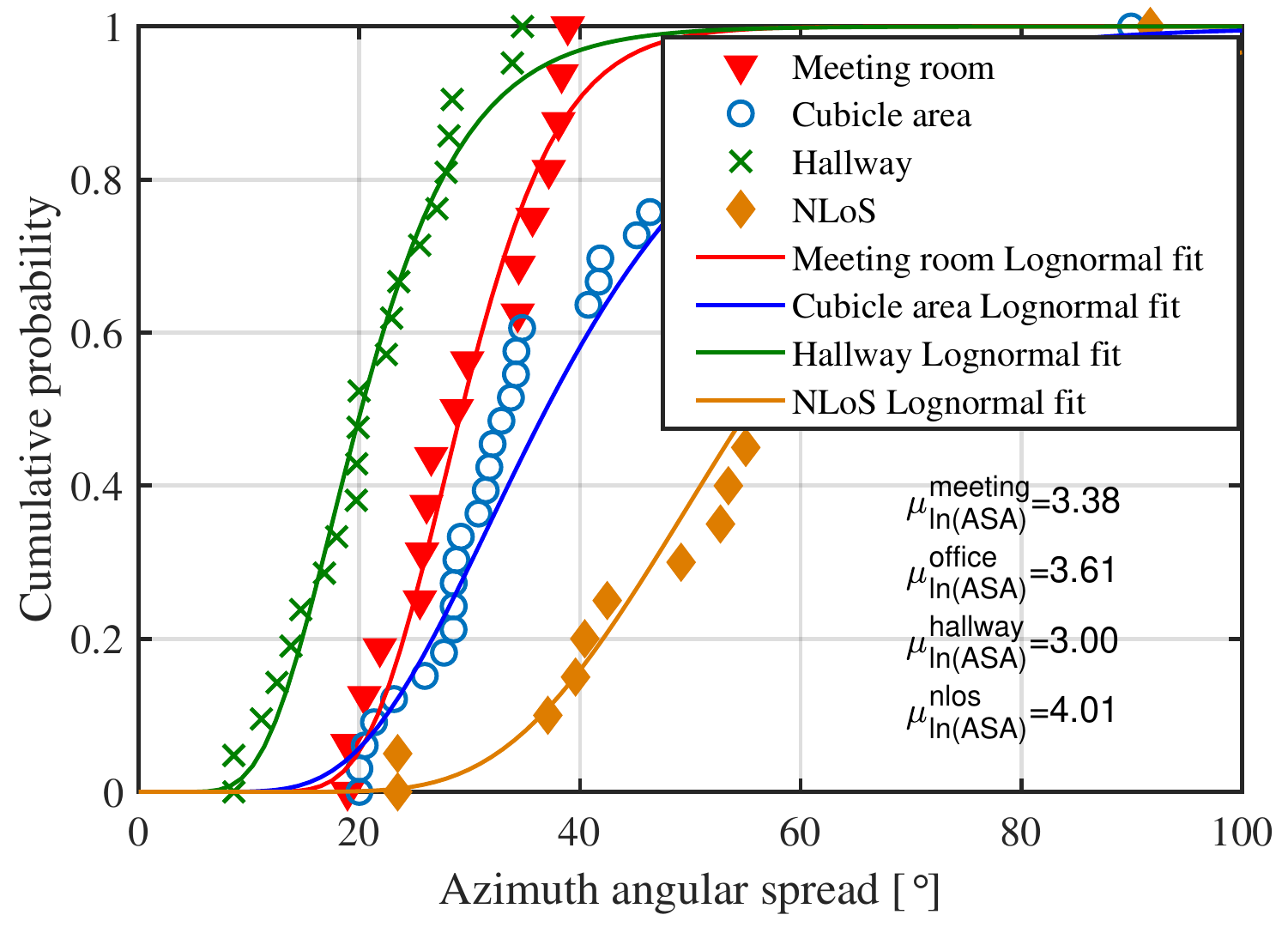}
\caption{Azimuth angular spread distribution of the measured channel in four scenarios at 201-209~GHz.}
\label{fig:asa_201-209}
\end{figure}

\subsection{Angular Spread}\label{sec:channel_asa}
Power dispersion in the angular domain is characterized by angular spread, which is important in characterizing angular properties for multiple-input-multiple-output (MIMO) channels. RMS angular spread with the $j^{\textrm{th}}$ reference direction is calculated as, 
    \begin{equation}
    \phi^j_{\text{rms}}=\sqrt{\frac{\sum^{N^{\text{MPC}}}_{i=1}(\text{MOD}(\phi_i+j\Delta\phi,360)-\bar\phi)^2 P^{\text{MPC}}_i}{\sum^{N}_{i=1}P^{\text{MPC}}_i}},
    \end{equation}
    with
    \begin{equation}
    \bar\phi=\frac{\sum^{N}_{i=1}\text{MOD}(\phi_i+j\Delta\phi,360) P^{\text{MPC}}_i}{\sum^{N}_{i=1}P^{\text{MPC}}_i},
    \end{equation}
where $\phi_i$ is the AoA of the $i^{\text{th}}$ MPC, $\text{MOD}(\cdot)$ denotes a modulo operation, $j\Delta\phi$ is an offset angle to adjust reference direction, and $P_i$ denotes the received power of the $i^{\text{th}}$ MPC. To eliminate the effect of the reference angle on the calculated RMS angular spread, we take $\phi_{\text{rms}}=\min_j({\phi^j_{\text{rms}}})$ as the RMS angular spread.
\par RMS azimuth angular of arrival spread (ASA) in the indoor scenarios at 201-209~GHz and the log-normal distribution fitting are shown in Fig.~\ref{fig:asa_201-209}. The average log values of RMS ASA in the meeting room, cubicle area, hallway, and NLoS case are, 3.38, 3.61, 3.00, and 4.01, respectively. The NLoS case shows the largest RMS ASA, while the hallway has the smallest RMS ASA. Hence, we find that RMS DS is positively related to RMS ASA, i.e., the larger RMS DS corresponds to the larger RMS ASA, by comparing RMS DS and RMS ASA in the four scenarios. This is reasonable as the power is likely to disperse in the angular domain if it spreads in the delay domain.

\begin{figure}
\centering
\includegraphics[width=0.6\textwidth]{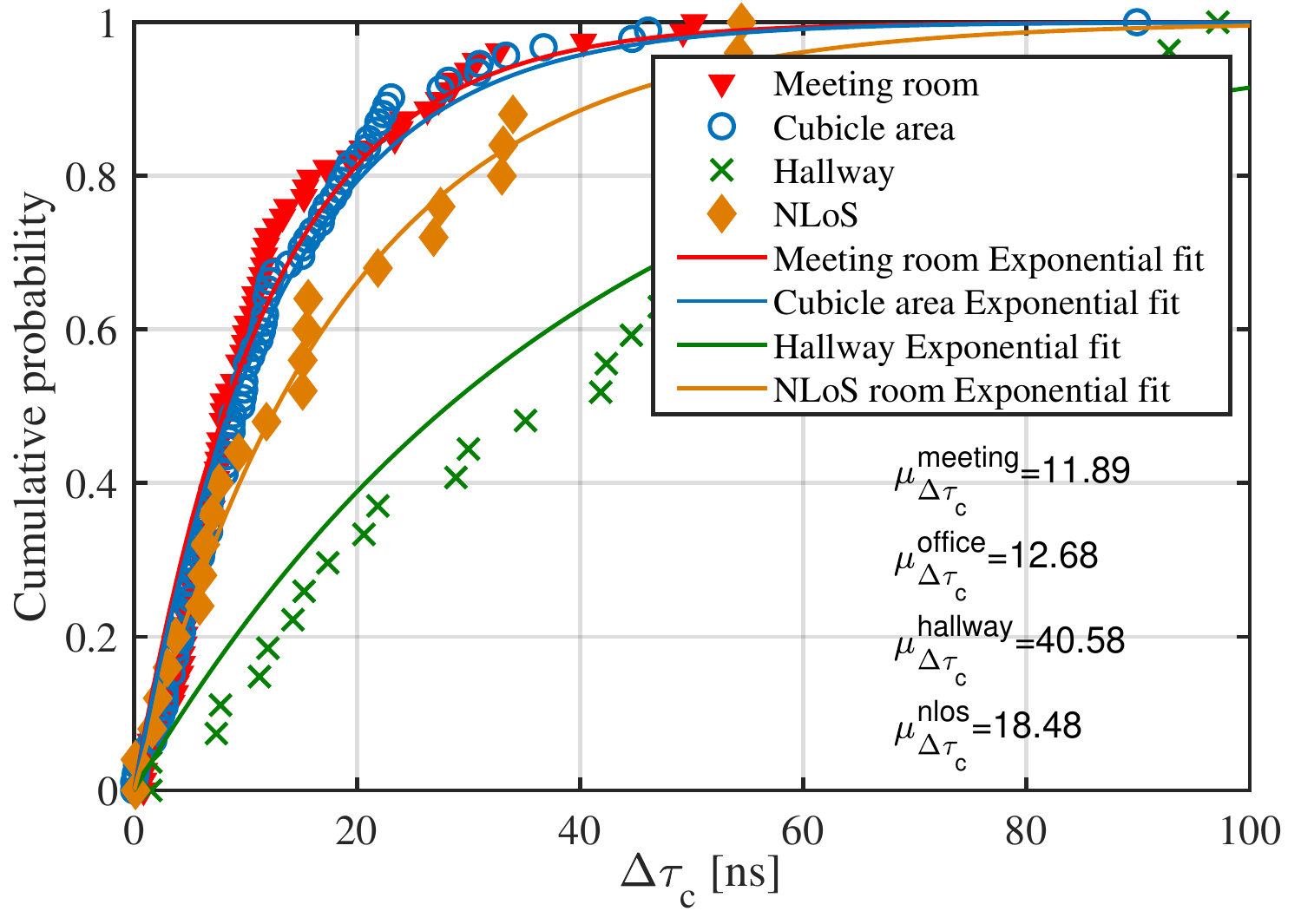}
\caption{Cluster delay difference distribution of the measured channel in four scenarios at 201-209~GHz.}
\label{fig:CDD_201-209}
\end{figure}

\subsection{Inter-cluster Delay}
In statistical channel models including the well-known Saleh-Valenzuela (S-V) model~\cite{saleh1987statistical} and NYU model~\cite{ju2021millimeter}, the arrival of clusters is often modeled by a Poisson process, i.e., the ToA difference between two adjacent clusters, $\Delta \tau_c$, is regarded to be exponentially distributed. Therefore, we use exponential distributions to fit $\Delta \tau_c$ in different scenarios at 201-209~GHz. As shown in Fig.~\ref{fig:CDD_201-209}, the exponential distribution fitting has a good agreement with the measured values in the four scenarios. The average values of $\Delta \tau_c$ in the meeting room, cubicle area, hallway, and NLoS case are 11.89~ns, 12.68~ns, 40.68~ns, and 18.48~ns, respectively. From the perspective of $\Delta \tau_c$, the meeting room and cubicle area show no significant difference. By contrast, the hallway scenario witnesses the largest $\Delta \tau_c$. The reason is that some reflection paths with shorter propagation distances combine with the LoS path into one cluster, which is the same as that for explaining the lower bound of reflection loss in the hallway scenario. The NLoS clusters that can be observed are those interacting with the cubicle area or the glass door. Those NLoS clusters have a longer propagation distance compared with the LoS cluster, as we discussed in Sec.~\ref{sec:MPC_LoSHallway}, which results in a longer inter-cluster delay.

\begin{figure}
\centering
\subfigure[Strongest reflection clusters.]{
\includegraphics[width=0.5\textwidth]{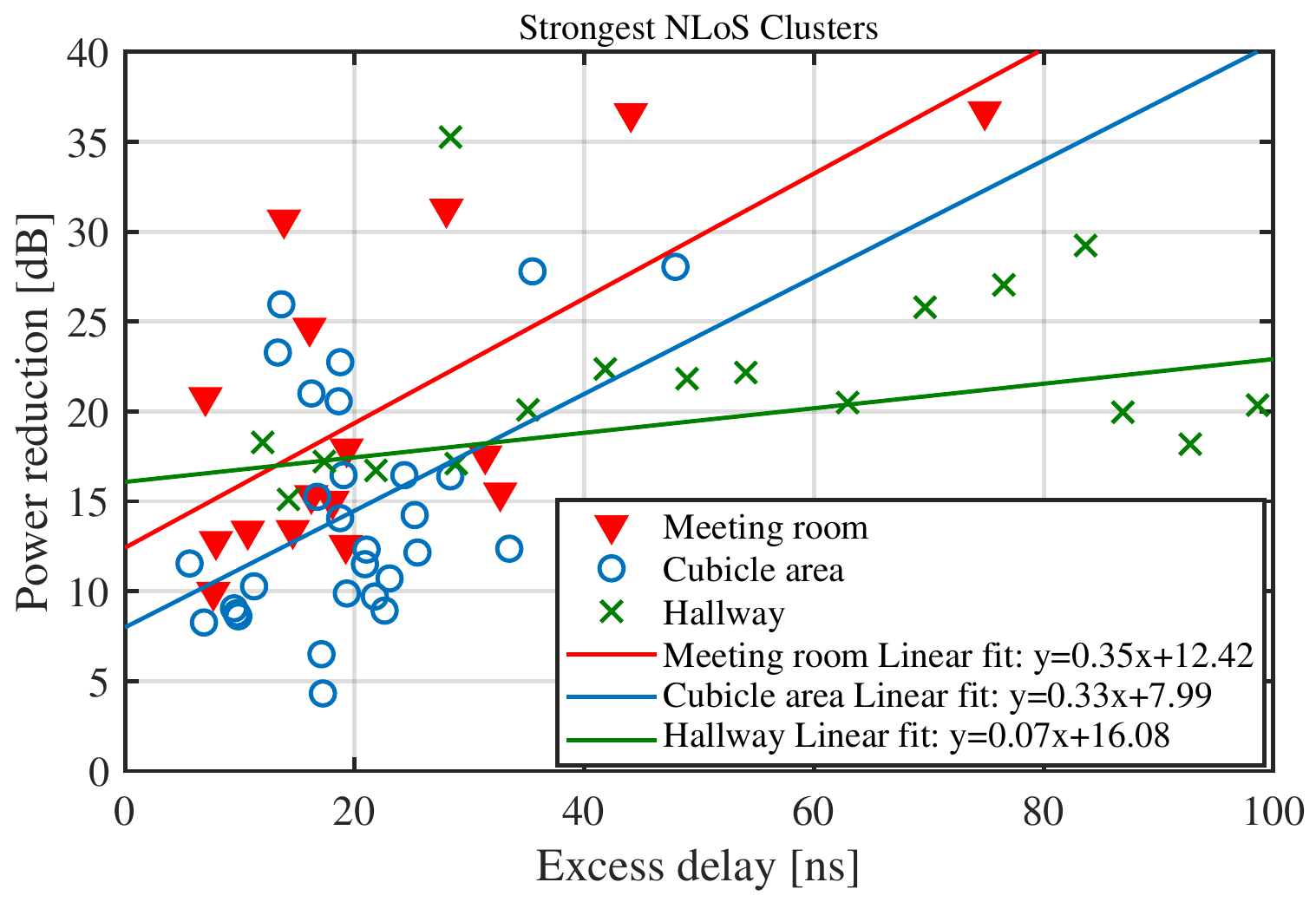}
    \label{fig:PR-strongest}
}

\subfigure[Second-strongest reflection clusters.]{
\includegraphics[width=0.5\textwidth]{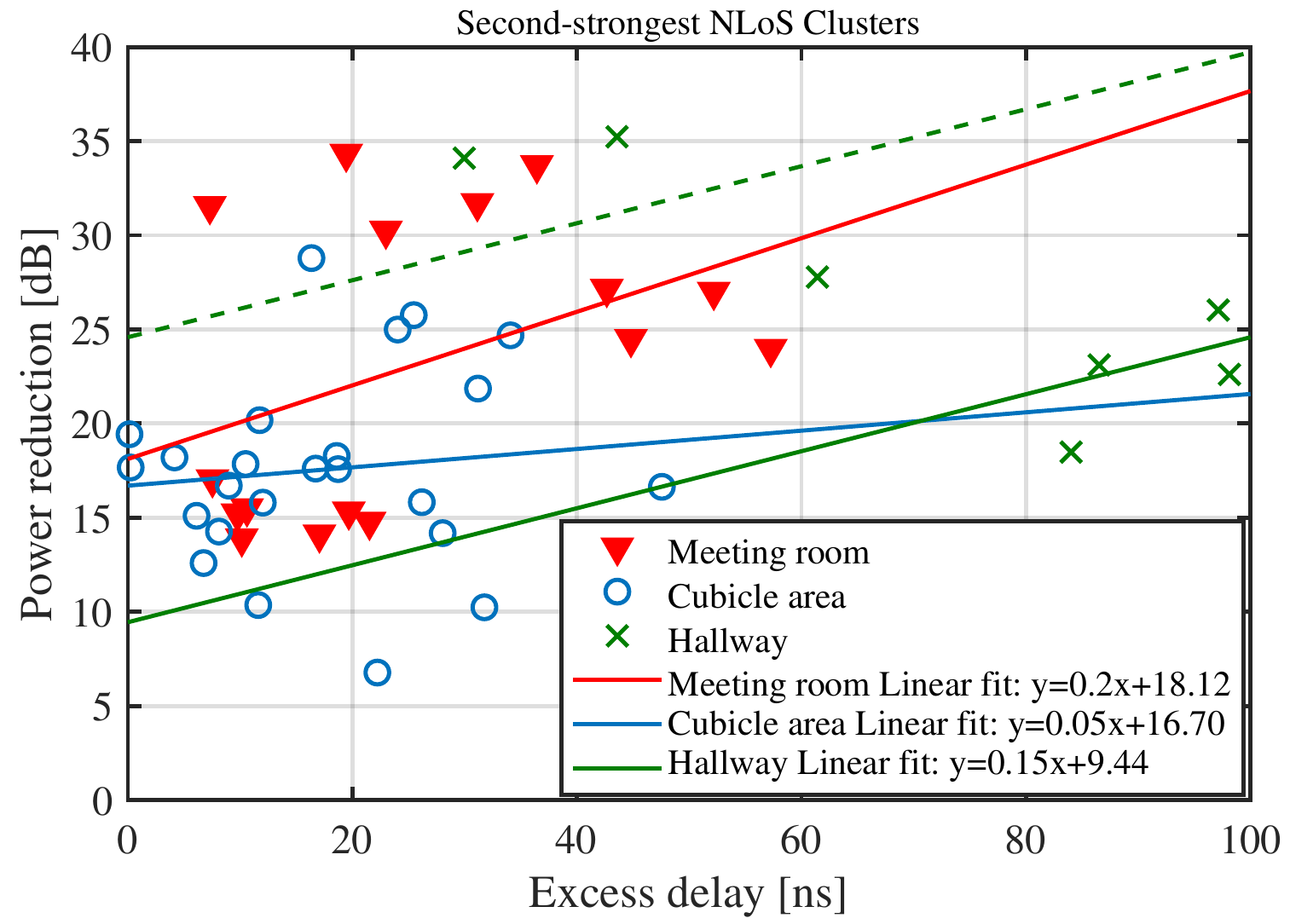}
\label{fig:PR-second-strongest}
}

\subfigure[Weak reflection clusters.]{
\includegraphics[width=0.5\textwidth]{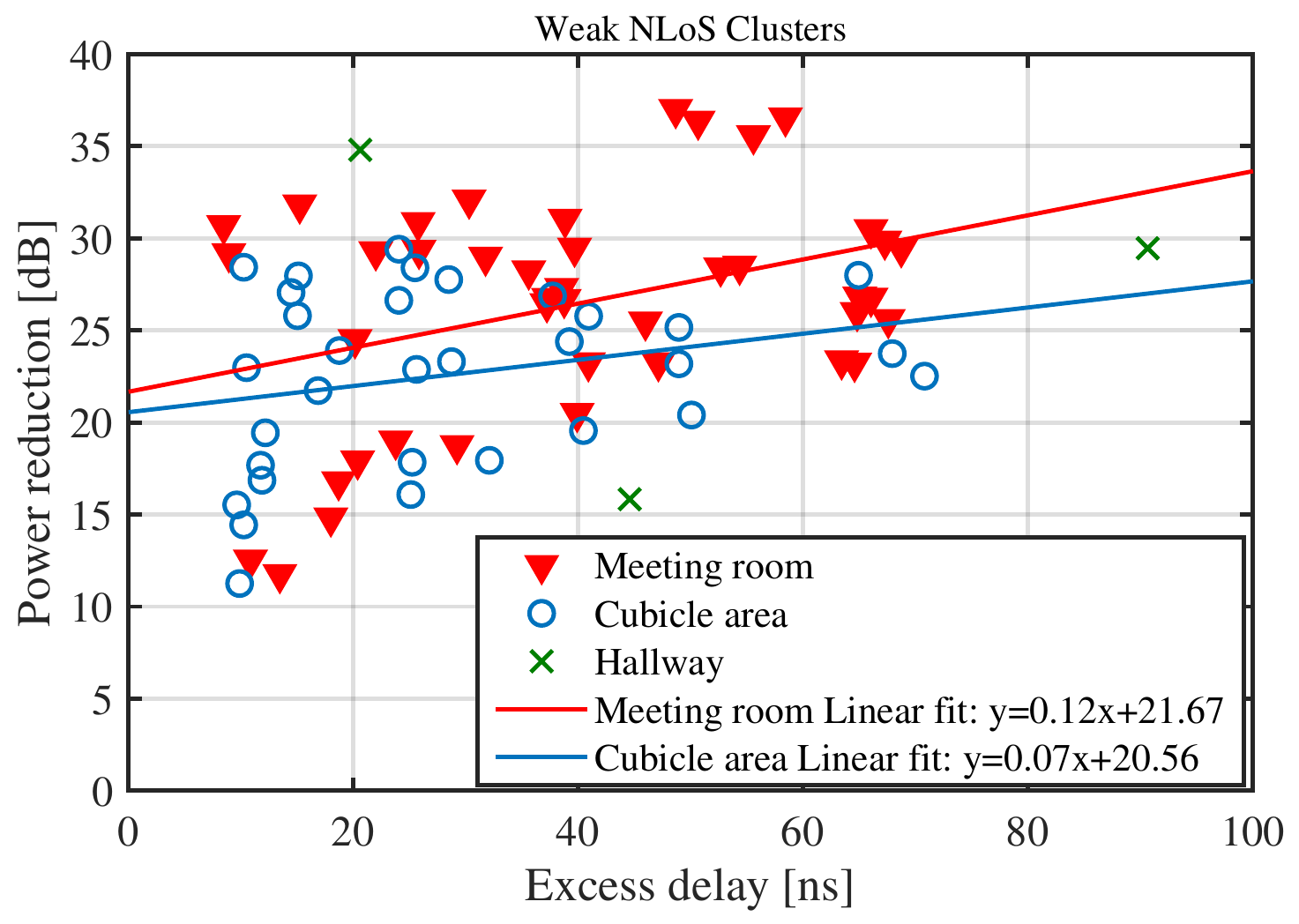}
\label{fig:PR-weak}
}
\caption{Normalized cluster loss versus excess delay for (a) strongest reflection clusters, (b) second-strongest reflection clusters, and (c) weak reflection clusters in four scenarios at 201-209~GHz.}
\label{fig:PR} 
\end{figure}
  
\subsection{Cluster Loss versus Excess Delay}
To analyze the relation between cluster power and excess delay, which draws much interest in channel modeling, we define \textit{normalized cluster loss} for the $i^{\text{th}}$ NLoS clusters, $\text{NCL}_i$, as the ratio between the LoS cluster power and the NLoS cluster power, as
\begin{equation}
    \text{CNL}_i \text{ [dB]}=10\log_{10}(\frac{P^c_{\text{LoS}}}{P_i^c}),
\end{equation}
where $P^c_{\text{LoS}}$ is the LoS cluster power. Similarly, the \textit{cluster excess delay} for a NLoS cluster is defined as the difference between the NLoS cluster delay and LoS cluster delay.
\par We further classify the NLoS clusters into three categories, i.e, the \textit{strongest} NLoS cluster, the \textit{second-strongest} NLoS cluster, and \textit{weak} NLoS cluster. The strongest NLoS cluster is the cluster with the highest received power among NLoS clusters in a Tx-Rx pair. The second-strongest NLoS cluster is only weaker than the strongest NLoS cluster. The other NLoS clusters are weak NLoS clusters. We depict the normalized cluster loss and the excess delay for the three kinds of NLoS clusters in Fig.~\ref{fig:PR}. In addition, we perform the linear fitting on the normalized cluster loss and excess delay for those NLoS clusters, which is a key assumption in the SV model. In Fig.~\ref{fig:PR-second-strongest}, the green dashed line denotes the linear fit in the hallway scenario with excess delay from 100 to 200~ns. The reason is that in the hallway scenario with long Tx-Rx distances, the ToA of some NLoS paths may exceed 100~ns and thus be wrapped to the range from 0 to 100~ns. For example, a cluster with ToA of 120~ns would appear at the position of 20~ns in the measured CIR. 
\par Generally, a larger excess delay leads to a higher normalized cluster loss, which is verified by the positive correlation coefficient. The power losses in the meeting room for these three kinds of NLoS clusters are higher than those in the cubicle area for the same excess delay, which results in smaller delay and angular spread in the meeting room as we discussed in Sec.~\ref{sec:channel_ds} and Sec.~\ref{sec:channel_asa}. This also verifies the relatively larger reflection loss in the meeting room as shown in Fig~\ref{fig:refl_loss}.

\section{Ray-tracing-statistical Hybrid Channel Models for Terahertz Indoor Propagation}
\label{Sec:ChannelModel}
 \par In this section, we propose a ray-tracing-statistical hybrid channel model for the indoor scenarios, in which the deterministic part of the channel model captures the dominate wall-reflection paths. 
 The hybrid channel impulse response, $h(\tau,\phi,f)$, consists of deterministic part $h_d(\tau,\phi,f)$ and statistical part $h_s(\tau,\phi,f)$, given as,
     \begin{equation}
    \label{eq:CIR}
    h(\tau,\phi,f)=h_d(\tau,\phi,f)+h_s(\tau,\phi,f).
    \end{equation}  
 \subsection{Determinsitc Channel Impulse Response}
 
 The deterministic channel impulse response, $h_d(\tau,\phi,f)$, is calculated as,
    \begin{equation}
    \label{eq:RT_CIR}
    h_d(\tau,\phi,f)=\sum^{L_{RT}}_{l=0}A_t(\vec{\theta_l})\alpha_{l}(f)\delta(\tau-\tau_{l})\delta(\phi-\phi_{l}),
    \end{equation}  
where $L_{RT}$ is the number of the traced paths. $A_t(\cdot)$ represents the antenna pattern at Tx. $\alpha_{l}(f)$, $\tau_{l}$, $\phi_{l}$ and $\vec{\theta}_{l}$ denote the amplitude gain, ToA, azimuth AoA and AoD vector of the $l^{\text{th}}$ traced path, respectively. In particular, the amplitude gain $\alpha_{l}(f)$ is calculated as,
    \begin{equation}
    \alpha_{l}(f)=\frac{|\Gamma|^{R_l}}{4\pi f\tau_{l}{(f)}}
    \end{equation}
    where $\Gamma$ is the reflection coefficient of the $l^{\mathrm{th}}$ RT cluster and statistically modeled in Sec.~\ref{sec:reflection-loss}, $R_l$ is the reflection times of the $l^{\mathrm{th}}$ RT cluster. 
  
  In this work, the deterministic traced rays are the multi-paths from the walls. As we discussed in Sec.~\ref{Sec:ChannelParameterCharacterizations}, the wall-reflection paths are not obvious and dominant in the cubicle area and NLoS case in the office room campaign. Therefore, still under the general hybrid model framework, the deterministic channel impulse response in these two scenarios are not considered.
 \subsection{Stasitical Channel Impulse Response}
 The statistical channel impulse response, $h_s(\tau,\phi,f)$, is calculated as,
    \begin{equation}
    \begin{split}
    h_s(\tau,\phi,f)=&\sum^{ N}_{n=1}\sum^{{M_n}}_{m=1}\alpha_{n,m}\delta(\tau-\tau_{n,m})\delta(\phi-\phi_{n,m}),
    \label{eq:statistical-cir}
    \end{split}
    \end{equation}
    where $N$ is the number clusters. $M_n$ denotes the number of subpaths in the $n^{\mathrm{th}}$ cluster. $\alpha_{n,m}$, $\tau_{n,m}$, $\phi_{n,m}$ denote the amplitude gain, ToA, azimuth AoA of the $m^{\mathrm{th}}$ subpath in the $n ^{\mathrm{th}}$ cluster, respectively. Next, we model these channel parameters of the statistical part in detail.

\subsubsection{Amplitude Gain}
The amplitude gain of the $m^\text{th}$ subpath in the $n^\text{th}$ cluster is calculated as,
\begin{equation}
    \alpha_{n,m}=\sqrt{\frac{P_{n,m}}{\text{PL}^\text{CI}}}
\end{equation}
where $P_{n,m}$ denotes the power of the $m^{\text{th}}$ subpath in the $n^{\text{th}}$ cluster, and $\text{PL}^\text{CI}$ is the calculated path loss in the linear scale based on the proposed CI model given in Sec. III.
    
	\subsubsection{Number of Clusters}
	The number of clusters $N$ is modeled by the Poisson distribution, given by,
	\begin{equation}
	    P(N=k)=\frac{\lambda_N^k}{k!}e^{-\lambda},\quad k=0,1,\dots,\infty
	\end{equation}
where $\lambda_N$ is the parameter given in Table~\ref{tab:table_conclude}. The number of subpaths in each cluster is set to be the average number of subpaths in a cluster from our channel measurements for each scenario
\begin{figure}[ht]
\centering
\subfigure[DS.]{
\includegraphics[width=0.45\textwidth]{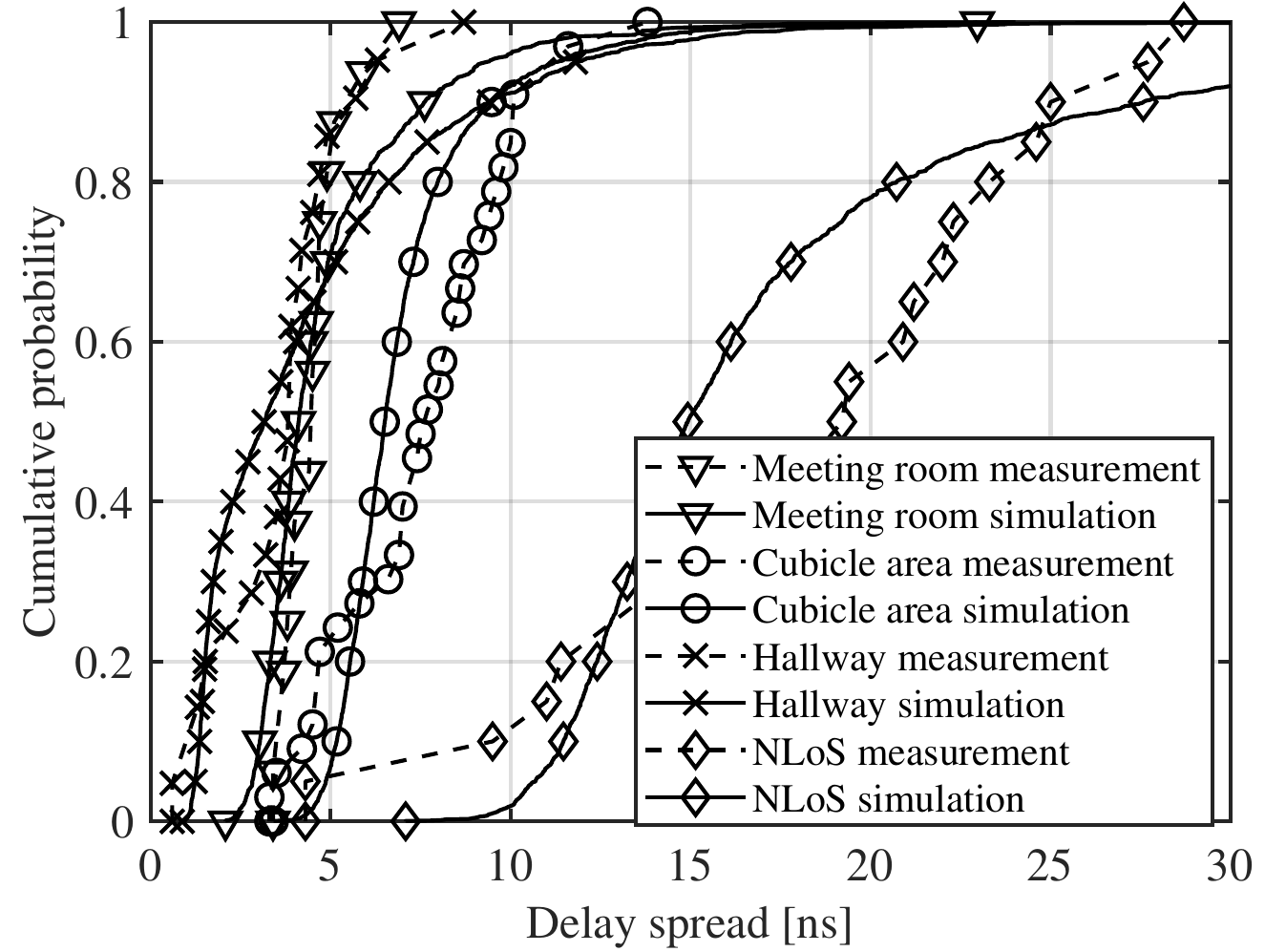}
\label{fig:DS_validation_220}
}
\centering
\subfigure[ASA.]{
\includegraphics[width=0.45\textwidth]{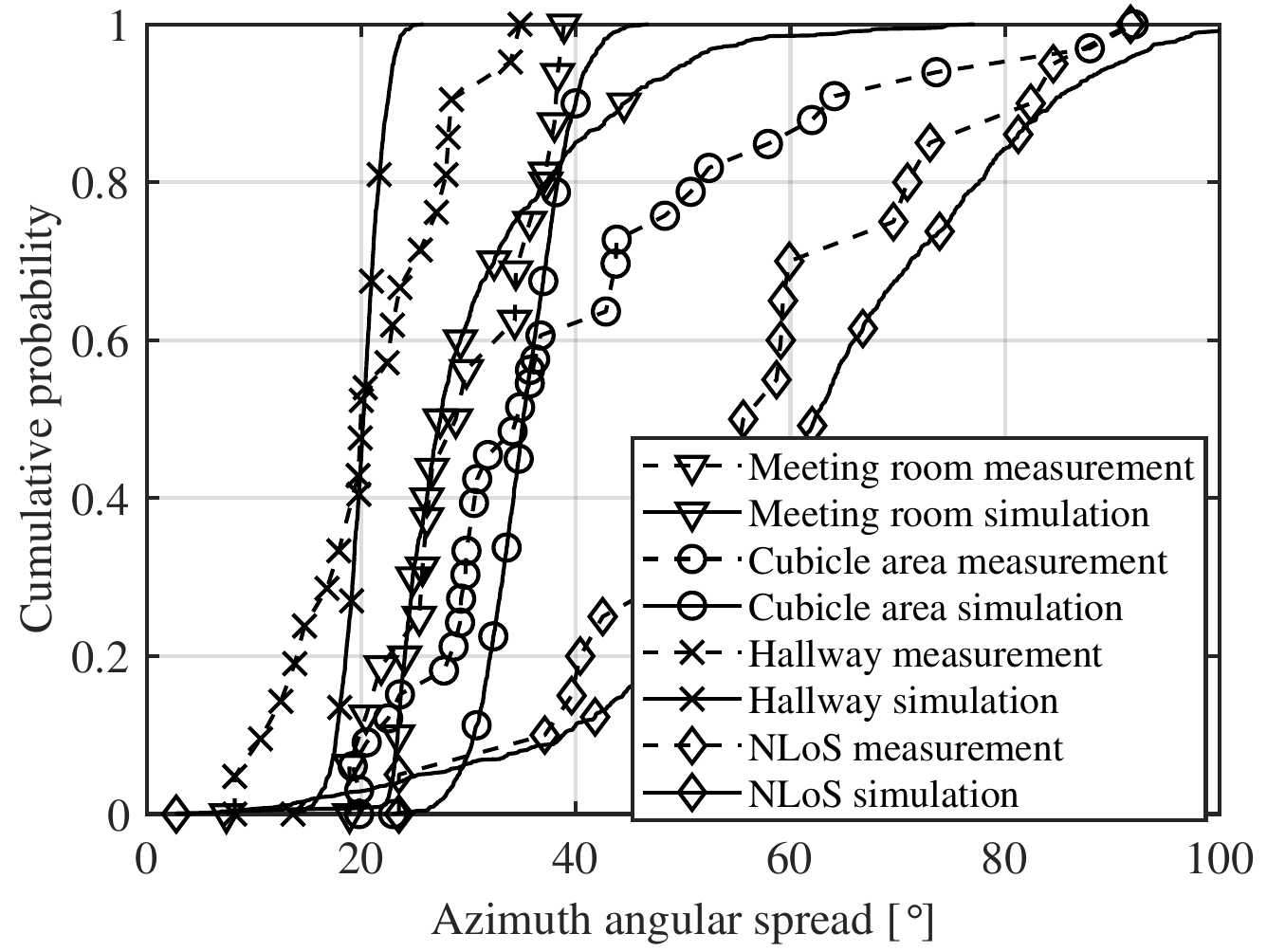}
\label{fig:AS_validation220}
}
\caption{DS and ASA validation of the proposed hybrid channel model with the measurement results for the four indoor scenarios at 201-209~GHz.}
\label{fig:model_validation_220}
\end{figure}
\subsubsection{Cluster Delays}
In our statistical channel model, the cluster delay is defined as the ToA of the $1^{\mathrm{th}}$ subpath in the $n ^{\mathrm{th}}$ cluster, i.e., $\tau_{n,1}$. The difference of cluster delay between two adjacent clusters is drawn randomly with the exponential delay distribution,
\begin{equation}
    \tau_{n+1,1}-\tau_{n,1}=-r_\tau\sigma_\tau \ln(X_n),
\end{equation}
where $r_\tau$ and $\sigma_\tau$ are two delay distribution proportionality factors, $X_n$ is uniformly distributed from 0 to 1. The product of two delay distribution proportionality factors, $r_\tau\sigma_\tau$, equals to the $\mu_{\Delta \tau_c}$ given in Table~\ref{tab:table_conclude}. The intra-cluster delays is also modeled as a Poisson process, given as,
\begin{equation}
    \tau_{n,m+1}-\tau_{n,m}=-r^c_\tau \ln(Y_n),
\end{equation}
where $r^c_\tau$ is an intra-cluster delay distribution proportionality factor, $Y_n$ is uniformly distributed from 0 to 1.
\subsubsection{Cluster Power}
The power of the non-LoS cluster is assumed to exponentially decrease with the cluster delay and calculated as,
\begin{equation}
    P^\prime_n=\exp(-\tau_{n,1}\frac{r_\tau-1}{r_\tau\sigma_\tau})\cdot 10^{(-\frac{Z_n}{10})}
\end{equation} 
where $Z_n\sim N(0,\xi^2)$ represents the per cluster shadowing fading effect. In the LoS case, the power of the first cluster is,
\begin{equation}
    P_{LoS}=\frac{K_r}{K_r+1}
\end{equation}
where $K_r$ is the K factor in the linear scale. The other cluster power is,
\begin{equation}
    P^c_n=\frac{1}{K_r+1}\frac{P^\prime_n}{\sum P^\prime_n}.
\end{equation}
\par In the NLoS case, the generated cluster power is normalized to 1,
\begin{equation}
    P^c_n=\frac{P^\prime_n}{\sum P^\prime_n}.
\end{equation}

\subsubsection{Angle of Arrival}
The azimuth AoA of a cluster, $\phi_{n,1}$, is modelled as wrapped Gaussian. The azimuth AoAs are obtained by the inverse Gaussian operation on the cluster power,
\begin{equation}
    \phi_{n,1}=C_n r_\phi\mu_ \mathrm{ASA}\sqrt{-\ln(P_n/\max(P_n))}
\end{equation}
where $C_n$ is a discrete random variable with uniform distribution to the set of $\{-1,1\}$, $r_\phi$ is an angular proportionality factor, and $\mu_{ASA}$ is the average measured ASA. The intra-cluster azimuth AoA is modeled as,
\begin{equation}
    \phi_{n,m}=D_n r^{c}_\phi\sqrt{-\ln(P_{n,m}/\max(P_{n,m}))}
\end{equation}
where $D_n$ is a discrete random variable with uniform distribution to the set of $\{-1,1\}$, $r^c_\phi$ is an angular proportionality factor for intra-cluster ASA, $P_{n,m}$ is the power of the $m^\text{th}$ subpath in the $n^\text{th}$ cluster.

\subsection{Model Validation}
The delay spread and azimuth angular spread are simulated by the proposed hybrid channel model for the four indoor scenarios, respectively. The cumulative probability functions of the simulated and measured values of delay spread and angular spread are compared in Fig.~\ref{fig:model_validation_220}. The average log values of simulated delay spread for the meeting room, cubicle area, hallway, and NLoS case are 1.50, 1.91, 1.18, and 2.79, which agree well with the measured ones: 1.50, 1.91, 1.20, and 2.83. The average log values of simulated azimuth angular spread are 3.39, 3.55, 2.99, and 4.06, and by comparison, the measured values are 3.38, 3.61, 3.00, and 4.01. Therefore, we conclude that the proposed statistical model is validated by the channel measurement results in the four indoor scenarios. 
\section{Conclusion} \label{Sec:Conclusion}
\begin{table*}
    \centering
    \caption{Summary of channel parameters in indoor environments at 201-209~GHz.}
    \includegraphics[width = \linewidth]{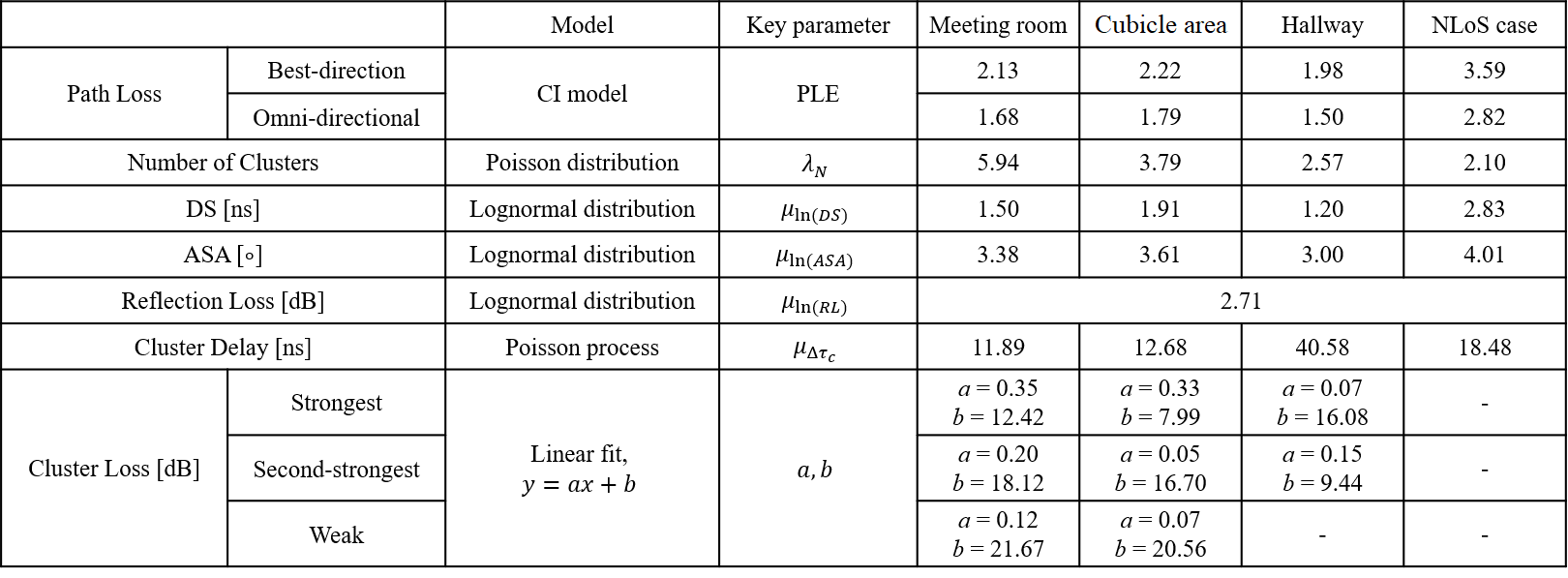}
    \label{tab:table_conclude}
\end{table*}
In this paper, we present our channel measurement campaigns in indoor scenarios at 201-209~GHz. Four different communication scenarios are covered, including the meeting room, cubicle area, hallway, and NLoS case. An antenna with HPBW of $60^\circ$ is used at Tx to guarantee wide coverage and Rx is equipped with a high-gain antenna and rotated for spatial scanning.
The best-direction and omni-directional path loss values in indoor scenarios are modeled by CI models. The LoS hallway scenario reports the lowest PLE due to the waveguide effect. The meeting room and cubicle area scenarios show similar large-scale fading as their path loss exponents are very close. 
Furthermore, the omni-directional path loss exponents are about 0.5 lower than the best-direction ones. Furthermore, we analyze the critical small-scale channel characteristics, including delay spread, angular spread, number of clusters, and cluster delay in each measured indoor scenario at 201-209~GHz, as summarized in Table~\ref{tab:table_conclude}.
\par Finally, we propose a general ray-tracing-statistical hybrid model is proposed and validated for the four indoor scenarios in the THz band. We conclude that even at the same frequency, the measurement results and analysis reveal that the channel characteristics in various indoor scenarios exhibit noticeable differences that need tailored parameter settings.
\bibliographystyle{IEEEtran}
\bibliography{IEEEabrv,CY_bib,bibliography.bib}
\vfill

\end{document}